\documentclass[10pt]{article}
\usepackage{amsmath, appendix, authblk, cleveref, color, mathtools, sectsty, subfig, txfonts}
\usepackage[top=1.5cm, bottom=1.5cm, left=2.5cm, right=2.5cm]{geometry}

\title{\Large Shape Optimization of Compliant Pressure Actuated Cellular Structures}
\author{Markus Pagitz and Remco I. Leine\\ \small markus.pagitz@inm.uni-stuttgart.de \large}
\affil{Institute for Nonlinear Mechanics, University of Stuttgart, 70569 Stuttgart, Germany}
\date{}

\begin{document}
    \maketitle

    \begin{abstract}
        Biologically inspired pressure actuated cellular structures can alter their shape through pressure variations. Previous work introduced a computational framework for pressure actuated cellular structures that is limited to two cell rows and central cell corner hinges. This article rigorously extends these results by taking into account an arbitrary number of cell rows, a more complicated cell kinematic that includes hinge eccentricities and varying side lengths as well as rotational and axial cell side springs. The nonlinear effects of arbitrary cell deformations are fully considered. Furthermore, the optimization is considerably improved by using a second-order approach. The presented framework enables the design of compliant pressure actuated cellular structures that can change their form from one shape to another within a set of one-dimensional $C^1$ continuous functions. Several examples are used to demonstrate the performance of the proposed framework.
    \end{abstract}


    \section{Introduction}
    \label{sec:Introduction}
\begin{figure}[htbp]
            \begin{center}
                \subfloat[]{
                    \includegraphics[height=0.34\textwidth]{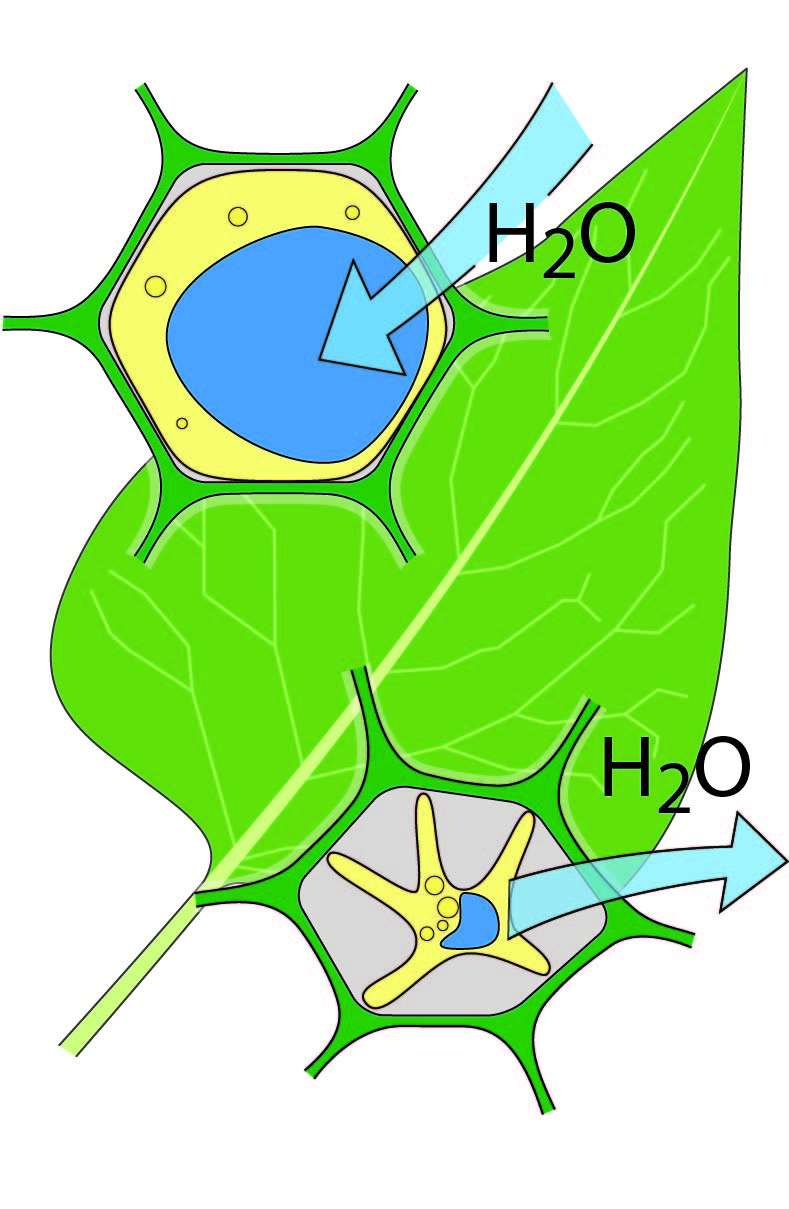}}\hspace{10mm}
                \subfloat[]{
                    \includegraphics[height=0.34\textwidth]{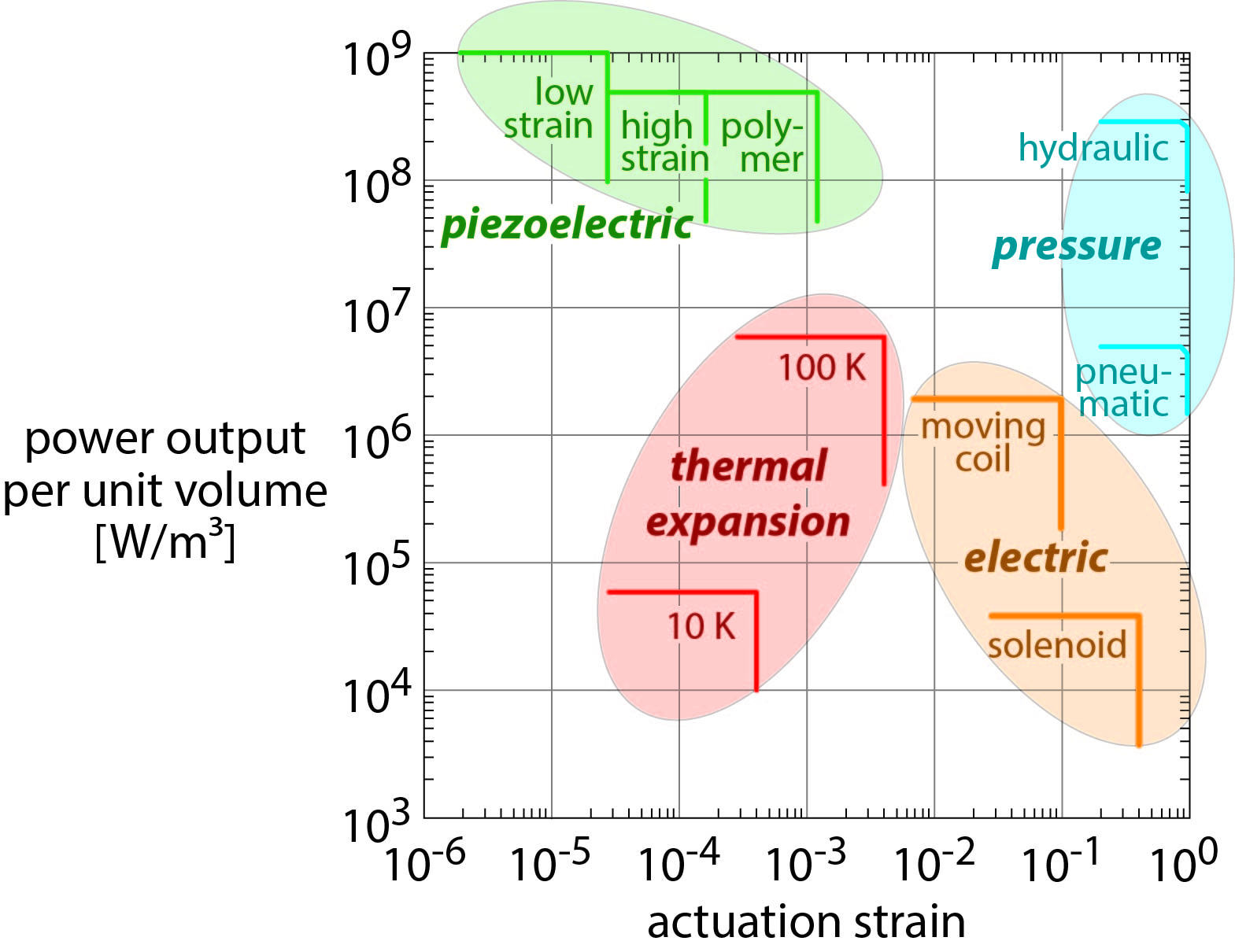}}
                \caption{(a) Osmotic hydration motors are used to vary cell pressures and thus to alter cell geometries. (b) Power output per unit volume versus strain of various actuation principles (data from Huber et al \cite{Huber1997}).}
              	\label{pic:Figure_1}
            \end{center}
        \end{figure}

        There exists a wide range of technologies that would immensely benefit from robust, strong, lightweight and energy efficient compliant structures that can change their form from one shape to another. For example, currently used aircraft slats are relatively heavy and the gaps between wings and slats increase noise levels, particulary during take off and landing. A comparison of existing actuation principles \cite{Huber1997} shows that pressure based actuators can generate large actuation strains and forces and thus have the largest potential to create such structures (Figure~\ref{pic:Figure_1}). Hence it is not surprising that the pressure driven nastic movement of plants attracted a lot of attention from various scientific communities during the last decade. A considerable research effort in this field was backed up by the Defense Advanced Research Agency, the National Science Foundation and the United States Army \cite{Wereley2012}. A comprehensive understanding of the nastic movement of plants requires various disciplines that range from biology and chemistry to material science and structural engineering. The focus of this article is on the structural engineering side. Therefore, no attention is given to the functionality of sub-cellular hydration motors \cite{Stahl2006} or plant cell materials \cite{Kerstens2001}. Instead, it is assumed that cellular structures are made from common engineering materials and that cell pressures are provided by an external source such as a compressor. Furthermore, only prismatic cells are considered. Hence, the problem reduces to the understanding of two-dimensional cell geometries and their interactions.\\

\begin{figure}[htbp]
            \begin{center}
                \begin{minipage}[c]{0.47\textwidth}
                    \subfloat[]{
                        \includegraphics[width=1\textwidth]{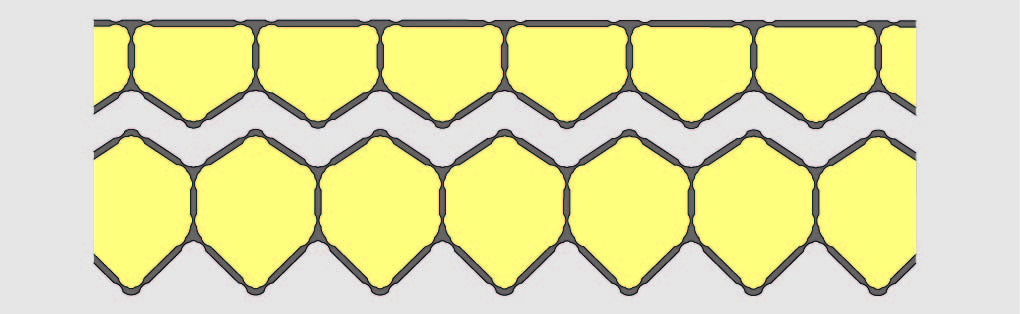}}

                    \subfloat[]{
                        \includegraphics[width=1\textwidth]{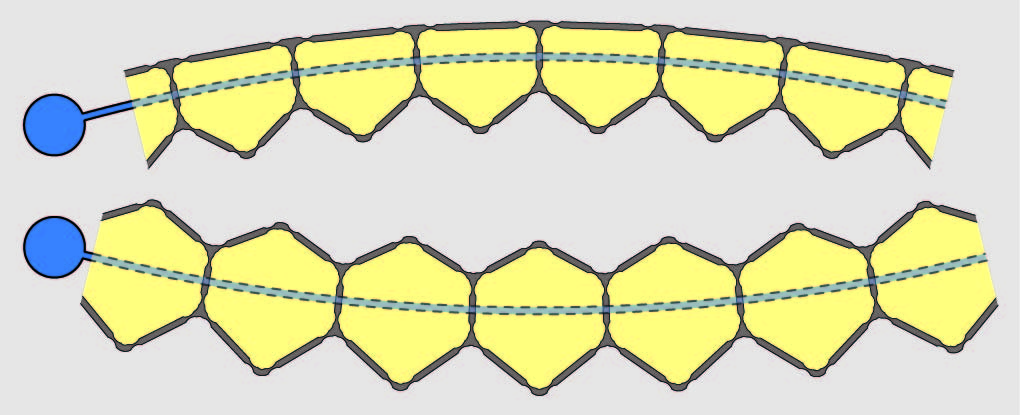}}

                    \subfloat[]{
                        \includegraphics[width=1\textwidth]{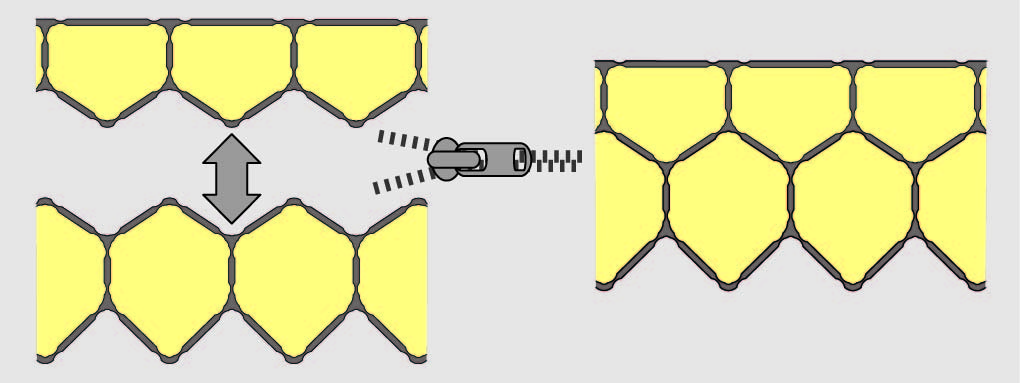}}

                    \subfloat[]{
                        \includegraphics[width=1\textwidth]{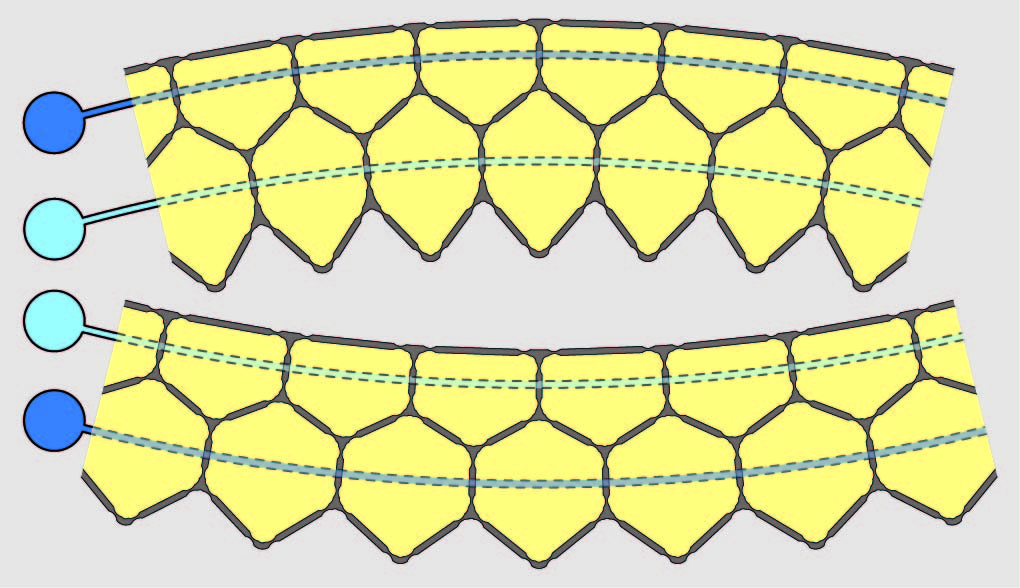}}

                    \subfloat[]{
                        \includegraphics[width=1\textwidth]{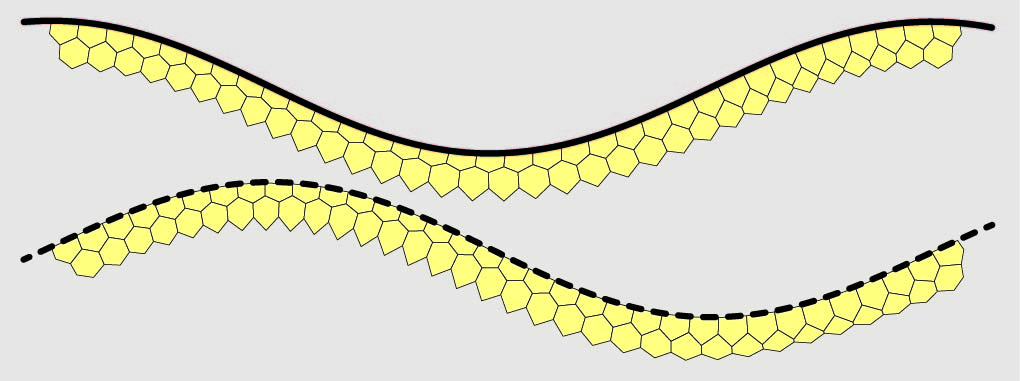}}

                    \subfloat[]{
                        \includegraphics[width=1\textwidth]{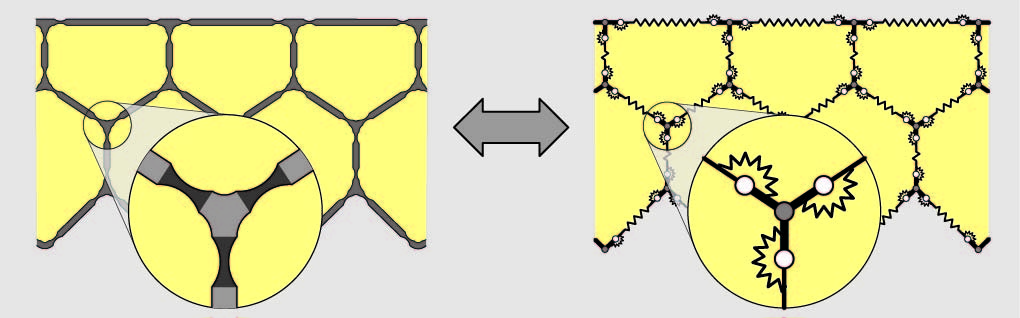}}
                \end{minipage}\hspace{5mm}
                \begin{minipage}[c]{0.45\textwidth}
                    \caption{
                    (a) Undeformed compliant cantilevers consist of a number of identical prismatic cells with pentagonal or hexagonal cross sections.
                    \vspace{6mm}\newline
                    (b) Upon pressurization, cantilevers deform into circular arcs. Corresponding radii depend on the material properties, cell geometry and  pressure.
                    \vspace{6mm}\newline
                    (c) Two cantilevers that are made from either pentagonal or hexagonal cells can be combined if the opposite cell geometries are compatible.
                    \vspace{6mm}\newline
                    (d) Equilibrium shape of a structure that is made from two connected cell rows can be altered by changing the pressure ratio between rows.
                    \vspace{6mm}\newline
                    (e) Cell geometries of a structure with $n_R$ cell rows can be optimized such that the structure changes its shape, for associated cell pressures, between $n_R$ given one-dimensional $C^1$ continuous functions.
                    \vspace{5mm}\newline
                    (f) A continuum and numerical model is used for the optimization of compliant pressure actuated cellular structures. The numerical model connects rigid cell corners via perfect hinges and rotational, axial springs. Both models are highly coupled. Altering the thickness of a single cell side changes the optimal geometry of all neighboring cell corners and sides. Furthermore, it affects the corresponding hinge eccentricities, axial springs and the $n_R$ equilibrium configurations of the numerical model. The latter leads to different maximum hinge and cell side stresses throughout the structure and thus, in return, alters the optimal geometry of all cell corners and sides.}
                    \vspace{6mm}
                    \begin{center}
                        \includegraphics[width=0.68\textwidth]{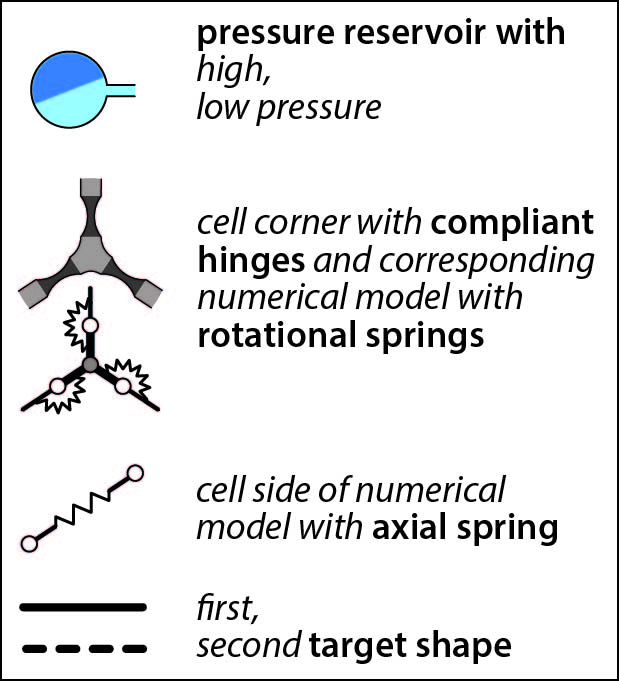}
                    \end{center}
                  	\label{pic:Figure_2}
                \end{minipage}
            \end{center}
        \end{figure}

        A brief overview of publications that investigate prismatic pressure actuated cellular structures is subsequently given. A concept based on plane symmetry groups was patented by Dittrich~\cite{Dittrich2005}. He combines convex and concave cells to form actuators that can replace double acting cylinders. A similar approach that uses pressurized and void cells was investigated by Luo and Tong~\cite{Tong2013}. Although not directly related to adaptive structures, Khire et al~\cite{Khire2006} studied inflatable structures that are made from a large number of uniformly pressurized hexagonal cells. Vos and Barret~\cite{Vos2010} subsequently patented a similar approach. Further work which investigates pressurized honeycombs can be found in \cite{Guiducci2014,Guiducci2015,Sun2014}. Numerical tools for the simulation and optimization of two-dimensional cellular structures were, among others, developed by Lv et al~\cite{Lv2013,Lv2014}. A concept for pressure actuated cellular structures that are made from separately pressurized rows of individually tailored prismatic cells (Figure~\ref{pic:Figure_2}) was patented by Pagitz et al~\cite{Pagitz2012-1,Pagitz2012-2}. It was shown in \cite{Pagitz2014-2} that these structures can be made from arbitrary engineering materials that range from elastomers to steel. Furthermore, it was shown in \cite{Pagitz2014-1} that their structural weight can be reduced and the overall stiffness increased with the help of cytoskeletons.\\

        Cell geometries can vary heavily throughout a compliant pressure actuated cellular structure (CPACS) so that it is hard if not impossible to use a homogenization based approach \cite{Davini2011} for the simulation and optimization. In contrast, equilibrium shapes of CPACS can be accurately computed by discretizing their cross sectional geometry with two-dimensional continuum finite elements (Figure~\ref{pic:Figure_2}). However, directly optimizing their geometry for given target shapes and cell pressures is impractical. This problem can be overcome by using a fully coupled continuum and numerical model. Such a separation is possible due to a concentration of bending strains in regions around cell corners that is mainly driven by the large axial cell side forces. The computation of cell corner geometries within the continuum model for given angles and side thicknesses is in itself a bilevel optimization problem and not treated in this article. This article focuses solely on the numerical model. Previously published framework \cite{Pagitz2012-1} is limited to structures with two cell rows and central cell corner hinges. The aim of this article is to extend the previous work by considering an arbitrary number of cell rows, the presence of hinge eccentricities and rotational as well as axial springs. Furthermore, the optimization is drastically improved by using a second order approach. This allows the design of CPACS that can change their shape between any given set of one-dimensional $C^1$ continuous functions.\\

        The outline of this article is as follows. Section~2 shows how CPACS can be advantageously abstracted by triangles, pentagons and cell sides. Furthermore, energy terms for these geometric primitives are given. The assembly of energy terms and the simulation, optimization of structures with an arbitrary number of cell rows is discussed in Section~3. Section~4 demonstrates the performance of the proposed framework with the help of several examples. Conclusions are given in Section~5.


    \section{Geometric Primitives}
\begin{figure}[htbp]
            \begin{center}
                \subfloat[]{
                    \label{pic:Figure_3a}
                    \includegraphics[height=0.35\textwidth]{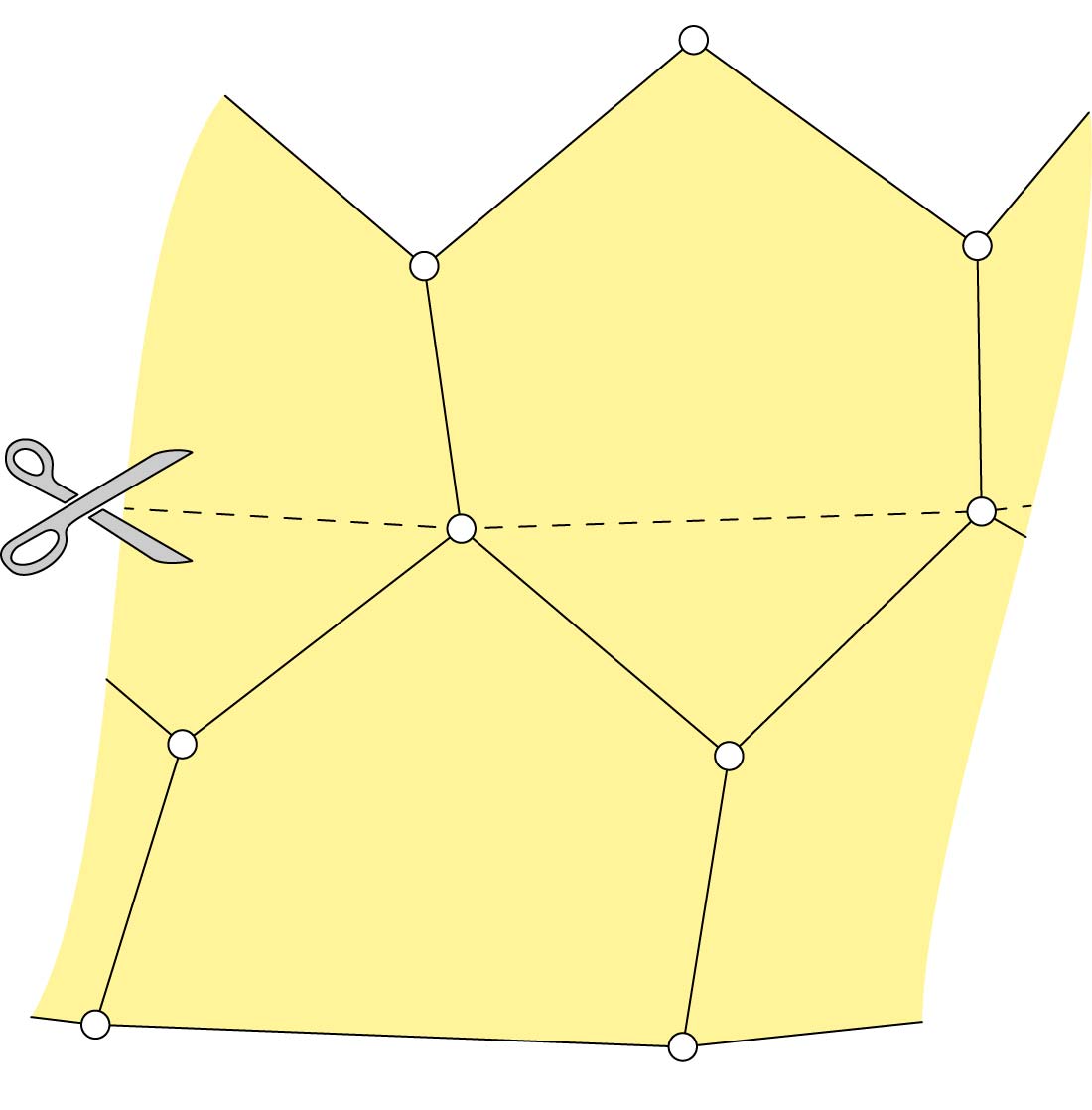}}\hspace{15mm}
                \subfloat[]{
                    \label{pic:Figure_3b}
                    \includegraphics[height=0.35\textwidth]{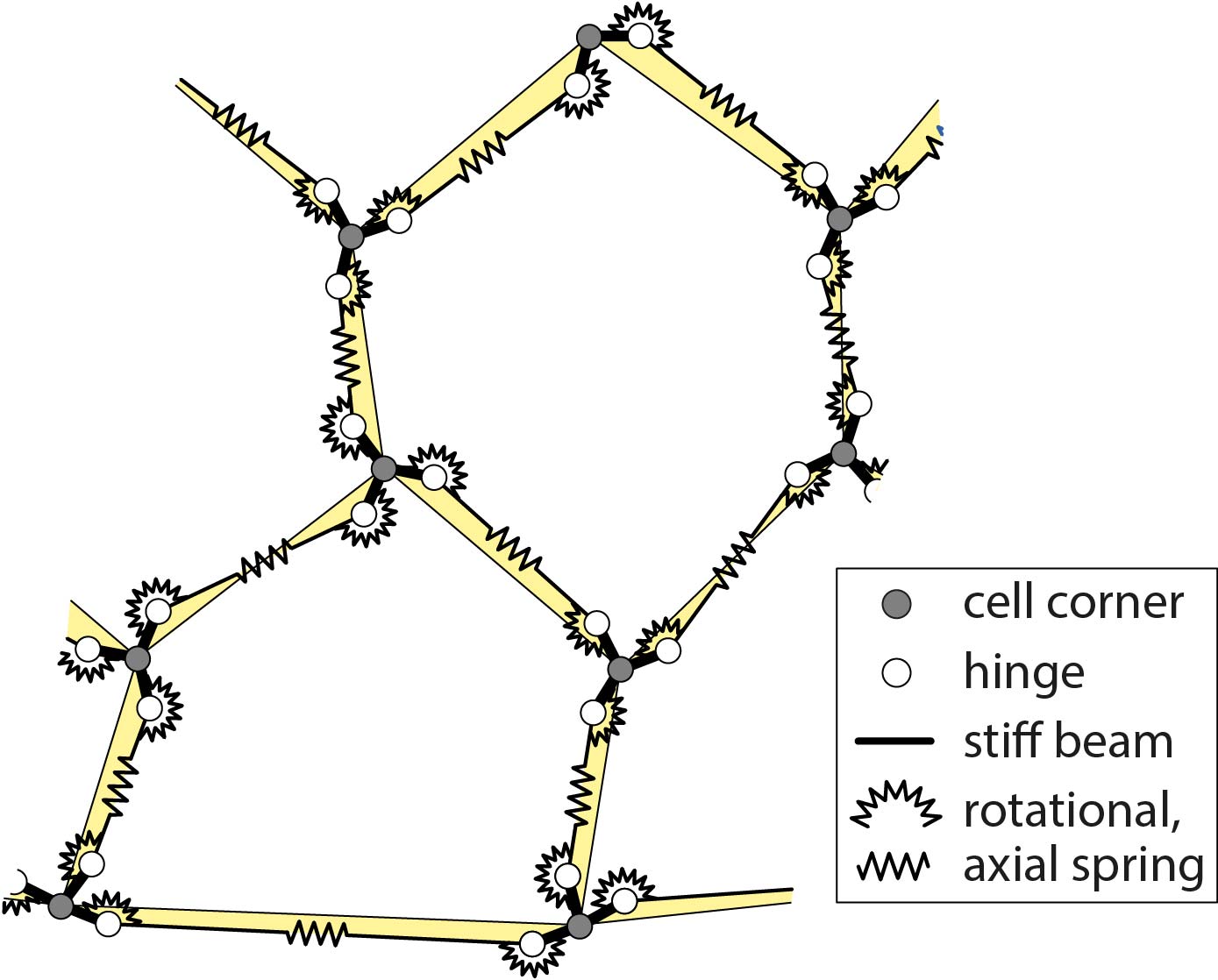}}
                \caption{Numerical model can be split into (a) pentagonal, hexagonal cells with central cell corner hinges and (b) cell sides with eccentric hinges and rotational, axial springs. Hexagonal cells can be divided into pentagonal and triangular subcells that are defined by neighboring pentagonal cells of the lower row.}
              	\label{pic:Figure_3}
            \end{center}
        \end{figure}

        The numerical model can be broken down into pentagonal and hexagonal cells that are bounded by cell sides (Figure~\ref{pic:Figure_3}). For computational reasons, each hexagonal cell is divided into a pentagonal and a triangular subcell. The latter is fully defined by its two neighboring pentagonal cells of the lower row. Cell side geometries are defined by hinge eccentricities, cell corner rotations and the distance between cell corners. Note that the area between a deformed side and the straight line between its cell corners is non-zero. This geometric reduction can be used as a basis for an object oriented implementation of the numerical framework.

        \subsection{Pentagonal Cells}
\begin{figure}[htbp]
                \begin{center}
                    \subfloat[]{
                        \includegraphics[height=0.2\textwidth]{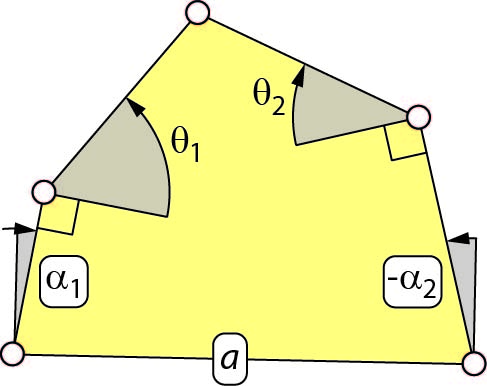}}\hspace{25mm}
                    \subfloat[]{
                        \includegraphics[height=0.2\textwidth]{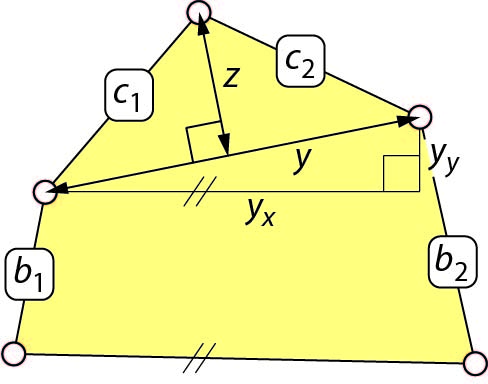}}
                    \caption{Variables of a pentagonal cell. (a) State variables $\mathbf{u}^\textrm{P}_{\alpha}$ and internal angles $\boldsymbol{\theta}$. (b) State variables $\mathbf{v}^\textrm{P}$ and internal lengths $y$, $z$.}
                    \label{pic:Figure_4}
                \end{center}
            \end{figure}

            The bottom row of CPACS consists of pentagonal cells. Furthermore, a pentagonal subcell forms the upper part of each hexagonal cell. A single pentagonal cell or subcell, shown in Figure~\ref{pic:Figure_4}, has five effective cell side lengths $a$, $b_1$, $b_2$, $c_1$ and $c_2$, which are defined as the distance between neighboring cell corners. The dependency between effective cell side lengths, hinge eccentricities and corner rotations is elaborated in Section~\ref{sec:Cell_Side}. The geometry of a pentagonal cell is further described by the external angles $\alpha_1$ and $\alpha_2$ and the internal angles $\theta_1$ and $\theta_2$. All effective cell side lengths and angles change with the pressurization of the system and therefore constitute the state of the pentagonal cell. It will prove to be advantageous to split the state variables into the groups
            \begin{align}
                \mathbf{u}^\textrm{P}_{\alpha} =
                \left[
                \begin{array}{ccc}
                    \alpha_1 & \alpha_2 & a
                \end{array}
                \right]^\top
                \hspace{5mm}\textrm{and}\hspace{5mm}
                \mathbf{v}^\textrm{P} =
                \left[
                \begin{array}{cccc}
                    b_1 & b_2 & c_1 & c_2
                \end{array}
                \right]^\top.
            \end{align}
            In the unpressurized and undeformed configuration, the vectors have the values $\mathbf{u}^\textrm{P}_{\alpha 0}$ and $\mathbf{v}^\textrm{P}_0$. The vector $\mathbf{u}^\textrm{P}_{\alpha 0}$ contains all design variables that are a priori chosen and usually not altered during the optimization process. The vector $\mathbf{v}^\textrm{P}_0$ contains all design variables that may be modified during the optimization process. The base side $a$ is a part of $\mathbf{u}^\textrm{P}_{\alpha}$ since it is an abstract term for pentagonal subcells that are a part of hexagonal cells. A superscript ``P" is used for pentagonal state variables. The length $y$ that divides the pentagon into a triangular and quadrilateral part is given by
            \begin{align}
                y = \sqrt{{y_x}^2 + {y_y}^2} =
                \sqrt{\left(a + \sin\left(\alpha_2\right)b_2 - \sin\left(\alpha_1\right)b_1\right)^2 + \left(\cos\left(\alpha_2\right)b_2 - \cos\left(\alpha_1\right)b_1\right)^2}
            \end{align}
            and the altitude $z$ can be expressed as
            \begin{align}
                z = \sqrt{{c_1}^2 - \frac{1}{4y^2}\left(y^2 + {c_1}^2 - {c_2}^2\right)^2}.
            \end{align}
            The internal angle $\theta_1$ of a pentagonal cell is completely determined by the state variables through
            \begin{align}
                \theta_1 &=
                \begin{cases}
                    \displaystyle\alpha_1 + \arcsin\left(\frac{y_y}{y}\right) + \arcsin\left(\frac{z}{c_1}\right)
                    & {c_2}^2 \leq y^2 + {c_1}^2\\
                    \displaystyle\alpha_1 + \arcsin\left(\frac{y_y}{y}\right) - \arcsin\left(\frac{z}{c_1}\right) + \pi
                    & {c_2}^2 > y^2 + {c_1}^2.
                \end{cases}
            \end{align}
            The expressions for the internal angle $\theta_2$ are derived in a similar manner. The previous equation could be written without a distinction of cases. However, this would result in lengthier expressions. The pressure potential of a pentagonal cell with central cell corner hinges and an internal pressure $p$ is
            \begin{align}
                \Pi^\textrm{P} = -p A^\textrm{P} =
                -\frac{p}{2} \left(\left(\cos\left(\alpha_1\right) b_1 + \cos\left(\alpha_2\right) b_2\right) a + \sin\left(\alpha_2 - \alpha_1\right) b_1 b_2 + y z\right),
            \end{align}
            where $A^\textrm{P}$ is the cross-sectional area. The gradients with respect to state variables $\mathbf{u}^\textrm{P}_{\alpha}$, $\mathbf{v}^\textrm{P}$ for the previous expressions can be found in \Crefrange{eqn:dy_duP}{eqn:dP_duP}.


        \subsection{Triangular Cells}
\begin{figure}[htbp]
                \begin{center}
                    \subfloat[]{
                        \includegraphics[width=0.4\textwidth]{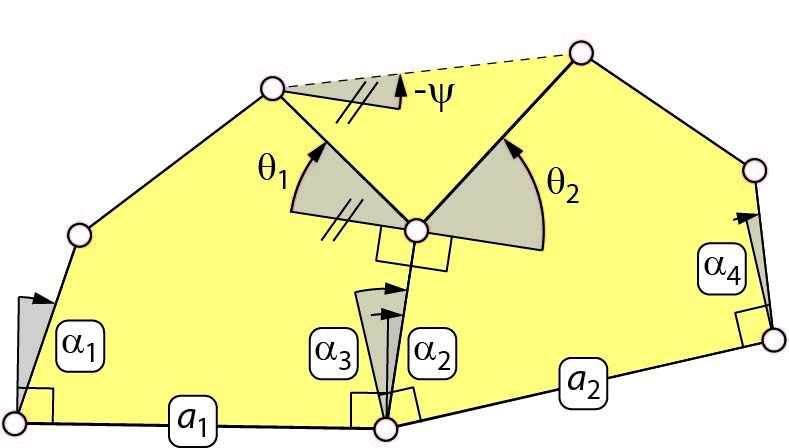}}\hspace{15mm}
                    \subfloat[]{
                        \includegraphics[width=0.4\textwidth]{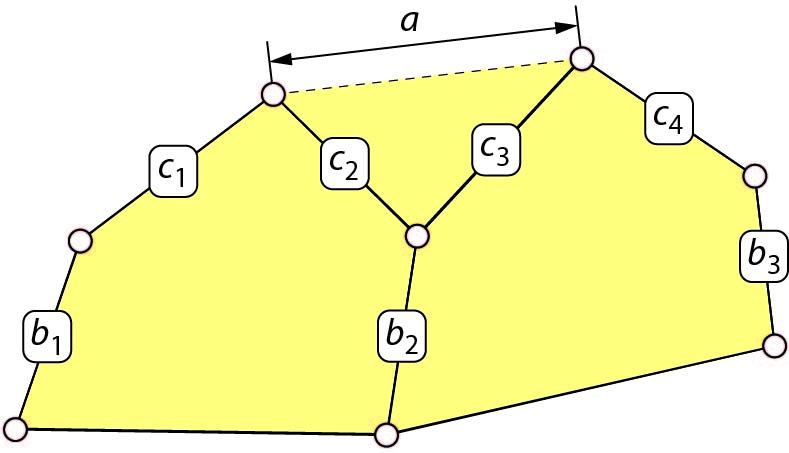}}
                    \caption{(a) State variables $\mathbf{u}^\textrm{T}_{\alpha}$, internal angles and (b) state variables $\mathbf{v}^\textrm{T}$, internal lengths of a triangular cell.}
                    \label{pic:Figure_5}
                \end{center}
            \end{figure}

            A triangular cell is defined by two neighboring pentagonal cells of the lower row. It constitutes together with a pentagonal cell in the upper row a hexagonal cell. Cell side lengths, abstract base lengths as well as state angles and internal angles of a triangular cell are shown in Figure~\ref{pic:Figure_5}. The state variables $\mathbf{u}^\textrm{T}_{\alpha}$  and cell sides $\mathbf{v}^\textrm{T}$ gather the kinematic quantities
            \begin{align}
                \mathbf{u}^\textrm{T}_{\alpha} =
                \left[
                \begin{array}{cccccc}
                    \alpha_1 & \alpha_2 & \alpha_3 & \alpha_4 & a_1 & a_2
                \end{array}
                \right]^\top
                \hspace{5mm}\textrm{and}\hspace{5mm}
                \mathbf{v}^\textrm{T} =
                \left[
                \begin{array}{ccccccc}
                    b_1 & b_2 & b_3 & c_1 & c_2 &  c_3 & c_4
                \end{array}
                \right]^\top,
            \end{align}
            where the superscript ``T" is used for state variables of a triangular cell. The abstract base side $a$ of the adjacent pentagonal cell is given by
            \begin{align}
                a = \sqrt{{c_2}^2 + {c_3}^2 + 2c_2 c_3 \cos\left(\theta_1 + \theta_2\right)}
            \end{align}
            and the internal angle $\psi$ of the triangular cell can be expressed as
            \begin{align}
                \psi = \theta_1 - \arccos\left(\frac{{a}^2+{c_2}^2-{c_3}^2}{2 a c_2}\right).
            \end{align}
            The pressure potential of the triangular cell without hinge eccentricities is determined by its area $A^\textrm{T}$ and reads as
            \begin{align}
                \Pi^\textrm{T} = -p A^\textrm{T} = -\frac{p}{2} c_2 c_3 \sin\left(\theta_1+\theta_2\right).
            \end{align}
            The gradients with respect to state variables $\mathbf{u}^\textrm{T}_{\alpha}$, $\mathbf{v}^\textrm{T}$ for the previous expressions can be found in \Crefrange{eqn:da_duT}{eqn:dP_duT}.


        \subsection{Cell Sides}
            \label{sec:Cell_Side}
\begin{figure}[htbp]
                \begin{center}
                    \subfloat[]{
                        \includegraphics[height=0.17\textwidth]{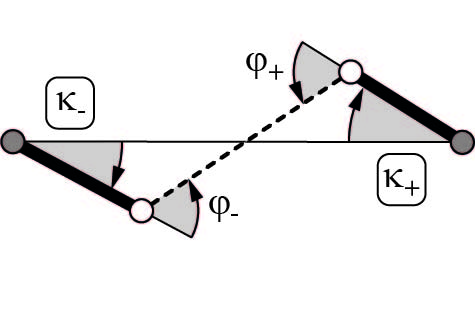}}\hspace{10mm}
                    \subfloat[]{
                        \includegraphics[height=0.17\textwidth]{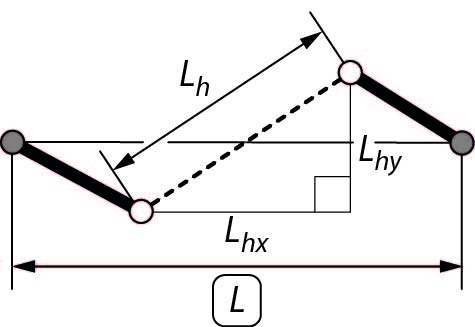}}\hspace{10mm}
                    \subfloat[]{
                        \includegraphics[height=0.17\textwidth]{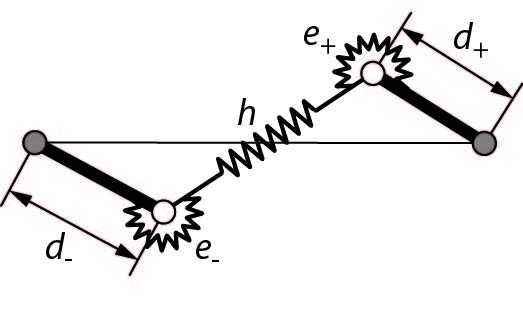}}
                    \caption{(a) State variables $\mathbf{u}_\kappa^\textrm{S}$ and bending angles $\varphi_\pm$. (b) State variable $v^\textrm{S}$ and length $L_h$. (c) Hinge eccentricities $d_\pm$ and rotational $e_\pm$, axial $h$ cell side springs.}
                    \label{pic:Figure_6}
                \end{center}
            \end{figure}

\begin{figure}[htbp]
                \begin{center}
                    \subfloat[]{
                        \includegraphics[width=0.28\textwidth]{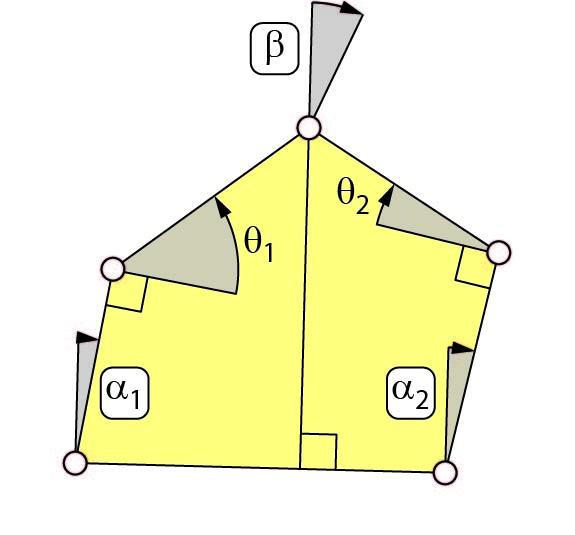}}\hspace{7mm}
                    \subfloat[]{
                        \includegraphics[width=0.28\textwidth]{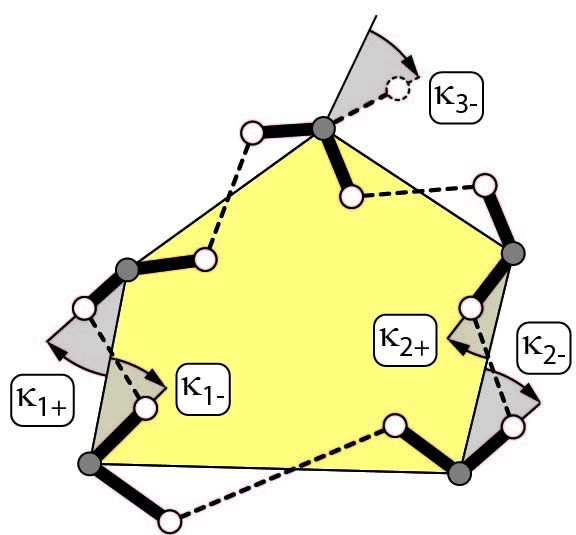}}\hspace{7mm}
                    \subfloat[]{
                        \includegraphics[width=0.28\textwidth]{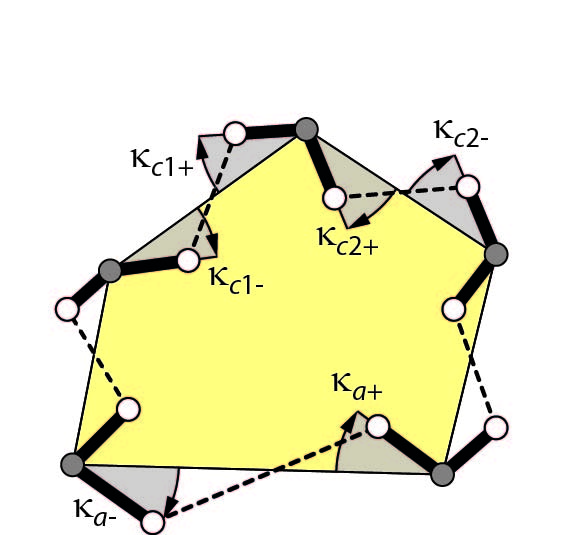}}
                    \caption{(a) State variables $\mathbf{u}^\textrm{P}_{\alpha}$, $\beta^\textrm{P}$ and internal angles of a pentagonal cell. (b) State variables $\mathbf{u}_{\kappa}^\textrm{S}$ for cell corner rotations are defined with respect to cell sides $b_1$ and $b_2$. (c) Derived state variables of cell sides $a$, $c_1$ and $c_2$.}
                    \label{pic:Figure_7}
                \end{center}
            \end{figure}

            The previously published numerical framework \cite{Pagitz2012-1} for pressure actuated cellular structures assumes rigid cell sides that are connected at cell corners via hinges. This assumption is valid as long as cell sides are relatively thin and stiff. In the following, it is outlined how rotational and axial springs as well as hinge eccentricities are taken into account. To simplify matters, it is assumed that undeformed cell sides are straight and that hinge eccentricities are invariant to cell side deformations. The latter assumption is valid since cell corners are usually compact and biaxially stressed. As before, the state variables are split into two parts. The first part
            \begin{align}
                \mathbf{u}^\textrm{S}_{\kappa} =
                \left[
                \begin{array}{cc}
                    \kappa_- & \kappa_+
                \end{array}
                \right]^\top
            \end{align}
            describes the angles between hinge eccentricities and the straight line that connects its neighboring cell corners. The second part
            \begin{align}
                v^\textrm{S} = L
            \end{align}
            is the distance between both cell corners, i.e.\ the effective cell side length. Further variables that are required to fully describe a cell side are hinge eccentricities $d_\pm$, rotational springs $e_\pm$ and an axial spring $h$ (Figure~\ref{pic:Figure_6}).
            The distance
            \begin{align}
                {L_h} = \sqrt{{L_{hx}}^2 + {L_{hy}}^2} =
                \sqrt{\left(L - \cos\left(\kappa_-\right) d_- - \cos\left(\kappa_+\right) d_+\right)^2 + \left(\sin\left(\kappa_-\right)d_- + \sin\left(\kappa_+\right)d_+\right)^2}.
            \end{align}
            between both cell side hinges is a function of state variables $\mathbf{u}_\kappa^\textrm{S}$, $v^\textrm{S}$ and hinge eccentricities $d_\pm$. It is possible to write the state variables $\mathbf{u}_{\kappa\, b}^\textrm{S}$ for cell sides $b_1$ and $b_2$ of a single pentagonal cell (Figure~\ref{pic:Figure_7}) as
            \begin{align}
                \mathbf{u}^\textrm{S}_{\kappa\, bj} =
                \left[
                    \begin{array}{cc}
                        \kappa_{bj-} &
                        \kappa_{bj+}
                    \end{array}
                \right]^\top =
                \left[
                \begin{array}{cc}
                    \kappa_{j-} &
                    \kappa_{j+}
                \end{array}
                \right]^\top,
            \end{align}
            where $j=1,2$. In contrast, state variables $\mathbf{u}_{\kappa\, a}^\textrm{S}$ and $\mathbf{u}_{\kappa\, c}^\textrm{S}$ of pentagonal cell sides $a$ and $c_1$, $c_2$ are a function of state variables $\mathbf{u}_{\kappa\, b}^\textrm{S}$ of cell sides $\mathbf{b}$ and pentagonal state variables $\mathbf{u}_\alpha^\textrm{P}$. Furthermore, they depend on an additional global state variable $\beta$ as shown in Figure~\ref{pic:Figure_7}
            \begin{align}
                \mathbf{u}^\textrm{S}_{\kappa\, a1} =
                \left[
                \begin{array}{c}
                \kappa_{1-} + \Delta\alpha_1\\
                \kappa_{2-} + \Delta\alpha_2
                \end{array}
                \right]
                \textrm{,}\hspace{5mm}
                \mathbf{u}^\textrm{S}_{\kappa\, c1} =
                \left[
                \begin{array}{l}
                    \kappa_{1+} + \Delta\theta_1\\
                    \kappa_{3-} + \Delta\theta_1 - \Delta\alpha_1 + \Delta\beta
                \end{array}\right]
                \hspace{5mm}\textrm{and}\hspace{5mm}
                \mathbf{u}^\textrm{S}_{\kappa\, c2} =
                \left[
                \begin{array}{l}
                    \kappa_{2+} - \Delta\theta_2\\
                    \kappa_{3-} - \Delta\theta_2 - \Delta\alpha_2 + \Delta\beta
                \end{array}\right].
            \end{align}
            For example, $\Delta \alpha = \alpha-\alpha_0$ is the difference between the current (pressurized) and reference (manufactured) configuration. It is assumed that $\Delta \beta=0$ for pentagonal cells that are located in the top, boundary cell row. This is due to the fact that global state variables $\beta$ are not required in the top row since there are no further pentagonal cells. Instead, they serve as the basis for variables $\kappa$ and thus can have any value, including zero. Bending angles of a single cell side are
            \begin{align}
                \varphi_- = \kappa_- + \arcsin\left(\frac{L_{hy}}{L_h}\right)
                \hspace{5mm}\textrm{and}\hspace{5mm}
                \varphi_+ = \kappa_+ + \arcsin\left(\frac{L_{hy}}{L_h}\right).
            \end{align}
            The pressure potential $\Pi^\textrm{S}_p$ of a cell side is the product of the differential pressure $\Delta p$ and the area between the deformed cell side and the straight line that connects its neighboring cell corners
            \begin{align}
                \Pi^\textrm{S}_p =
                - \frac{\Delta p}{2}
                \left(
                \sin\left(\kappa_+\right) \cos\left(\kappa_+\right) {d_+}^2 -
                \sin\left(\kappa_-\right) \cos\left(\kappa_-\right) {d_-}^2 +
                \left(
                \sin\left(\kappa_+\right) d_+ - \sin\left(\kappa_-\right) d_-
                \right) L_{hx}
                \right).
            \end{align}
            The strain potential $\Pi^\textrm{S}_e$ of a cell side consists of the rotational and axial strain energy, i.e.
            \begin{align}
                \Pi^\textrm{S}_e =
                \frac{1}{2} \left( e_- {\varphi_-}^2 + e_+ {\varphi_+}^2 + h {\Delta L_h}^2 \right),
            \end{align}
            where $\Delta L_h = L_h - L_{h0}$. Recall that $\Delta \varphi = \varphi$ since undeformed cell sides are straight. The total energy of a cell side is the sum of the pressure and strain energy
            \begin{align}
                \Pi^\textrm{S} = \Pi^\textrm{S}_p + \Pi^\textrm{S}_e.
            \end{align}
            The gradients of previous expressions with respect to cell side state variables $\mathbf{u}_{\kappa}^\textrm{S}$, $v^\textrm{S}$ and hinge eccentricities $\mathbf{w}^\textrm{S}$ as well as pentagonal state variables $\mathbf{u}_{\alpha}^\textrm{P}$, $\mathbf{v}^\textrm{P}$ can be found in \Crefrange{eqn:dL_duS}{eqn:dP_duS}.


    \section{Cellular Structure}
        \subsection{Variables}
\begin{figure}[htbp]
                \begin{center}
                    \subfloat[]{
                        \includegraphics[height=0.33\textwidth]{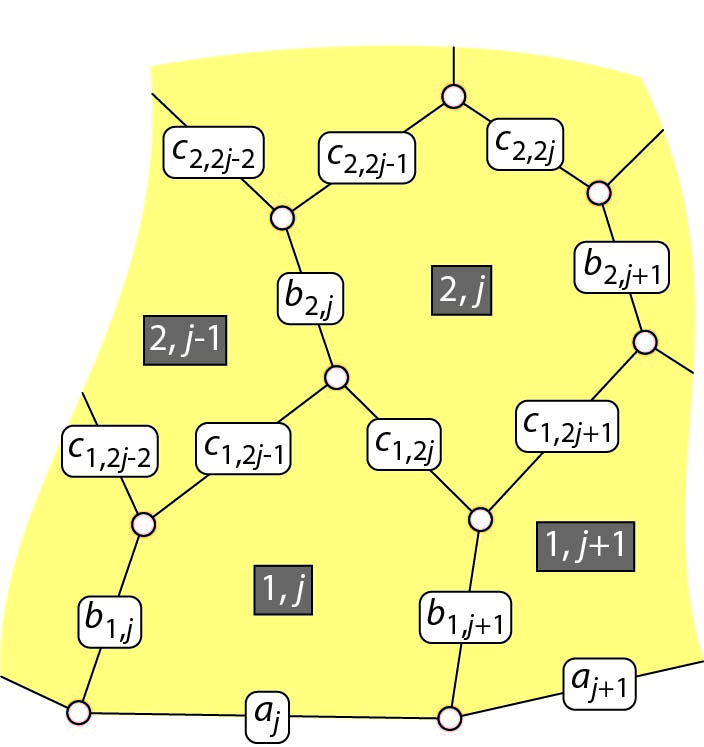}}\hspace{1mm}
                    \subfloat[]{
                        \includegraphics[height=0.33\textwidth]{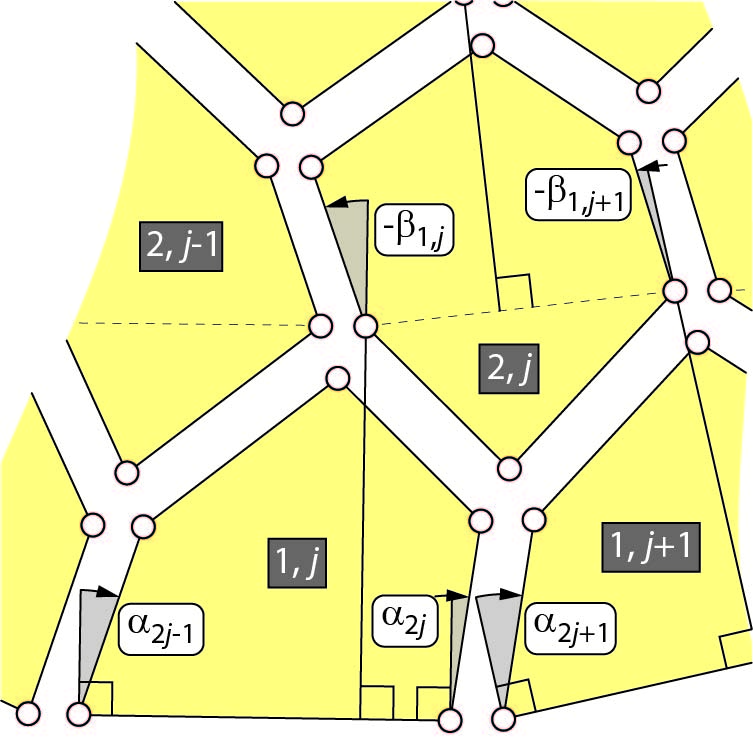}}\hspace{1mm}
                    \subfloat[]{
                        \includegraphics[height=0.33\textwidth]{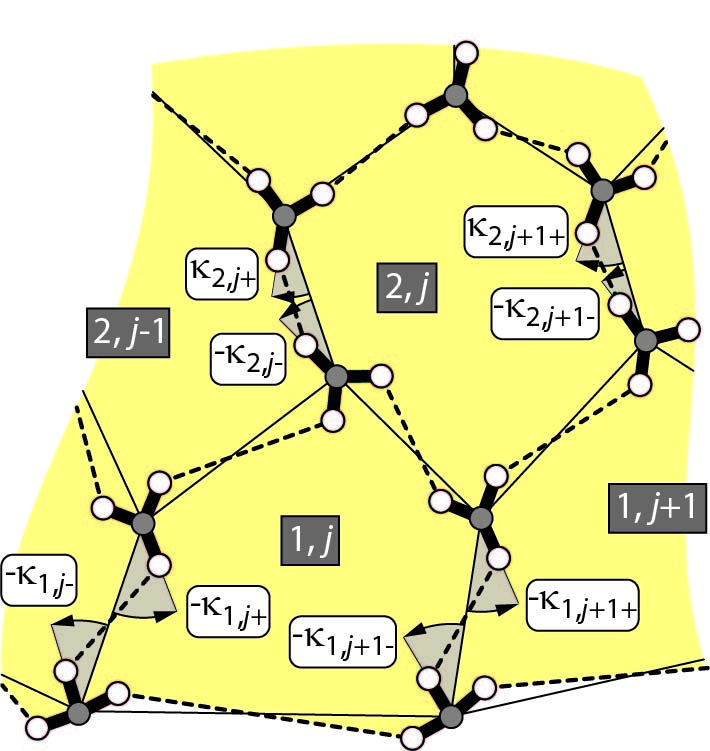}}
                    \caption{(a) Effective side lengths $\mathbf{v}$ and (b) state variables $\mathbf{u}_\alpha$ of a cellular structure with central cell corner hinges. (c) Cell corner rotations $\mathbf{u}_\kappa$ for hinge eccentricities.}
                    \label{pic:Figure_8}
                \end{center}
            \end{figure}

            The used notation for state variables, hinge eccentricities and internal angles, lengths of a cellular structure is summarized in Figure~\ref{pic:Figure_8}. The effective cell side lengths $\mathbf{v}$ between cell corners are
            \begin{align}
                \mathbf{v} =
                \left[
                \begin{array}{cccccc}
                    {\mathbf{b}_1}^\top &
                    {\mathbf{c}_1}^\top & \ldots &
                    {\mathbf{b}_{nR}}^\top &
                    {\mathbf{c}_{nR}}^\top &
                    {\mathbf{a}}^\top
                \end{array}
                \right]^\top\in\mathbb{R}^{nv}
            \end{align}
            where, for example $\mathbf{b}_i = \left[b_{i,1} \ldots b_{i,nP+2-i}\right]^\top$. Herein, $n_P$ denotes the number of base pentagons and $n_R$ is the number of cell rows. It can be seen that $\mathbf{v}$ incorporates, in contrast to $\mathbf{v}^\textrm{P}$ and $\mathbf{v}^\textrm{T}$, the non-abstract pentagonal base sides $\mathbf{a}$. The total number of cells $n_C$ and cell sides $n_v$ of a cellular structure are
            \begin{align}
                n_C = \frac{n_R}{2}\left(2n_P-n_R+1\right)
                \hspace{5mm}\textrm{and}\hspace{5mm}
                n_v = 3 n_C + n_P + n_R.
            \end{align}
            It is subsequently assumed that each additional cell row contains one cell less than the previous row. This is not a limitation of the proposed framework since arbitrary topologies at both ends can be modeled with the help of constraints. For example, adaptive modules as introduced in \cite{Pagitz2014-1} require boundaries at both ends whose geometries are invariant to cell pressures. This can be enforced by constraining the corresponding state variables. State variables $\mathbf{u}_\alpha$ of a cellular structure without hinge eccentricities are
            \begin{align}
                \mathbf{u}_{\alpha} =
                \left[
                \begin{array}{cccccc}
                    \alpha_1    & \ldots & \alpha_{2 nP} &
                    \beta_{1,1} & \ldots & \beta_{nR-1,nP-nR+2}
                \end{array}
                \right]^\top\in\mathbb{R}^{n\alpha}
            \end{align}
            where the number of state variables $n_\alpha$ is
            \begin{align}
                n_\alpha = n_C + n_P + n_R - 1.
            \end{align}
            Cellular structures with non-zero hinge eccentricities require additional state variables $\mathbf{u}_\kappa$ that describe cell corner rotations. They are expressed with respect to cell sides $\mathbf{b}$ so that
            \begin{align}
                \mathbf{u}_{\kappa}=
                \left[
                \begin{array}{ccc}
                    {\boldsymbol{\kappa}_1}^\top &
                    \ldots &
                    {\boldsymbol{\kappa}_{nR+1}}^\top
                \end{array}
                \right]^\top\in\mathbb{R}^{n\kappa}
                \hspace{5mm}\textrm{where}\hspace{5mm}
                \boldsymbol{\kappa}_i =
                \begin{cases}
                    \left[
                    \begin{array}{ccccc}
                        \kappa_{i,1-} & \kappa_{i,1+} & \ldots & \kappa_{i,nP-i+2-} & \kappa_{i,nP-i+2+}
                    \end{array}
                    \right]^\top & i \leq n_R\\
                    \left[
                    \begin{array}{ccc}
                        \kappa_{nR+1,1-} & \ldots & \kappa_{nR+1,nP-nR+1-}
                    \end{array}
                    \right]^\top & i > n_R
                \end{cases}.
            \end{align}
            The number $n_\kappa$ of state variables $\mathbf{u}_\kappa$ equals
            \begin{align}
                n_\kappa = 2n_C + n_P + n_R + 1.
            \end{align}
            Therefore, the state variables of a cellular structure with hinge eccentricities can be written for the reference (manufactured) and current (pressurized) configuration as
            \begin{align}
                \mathbf{u}_0 = \left[
                \begin{array}{cc}
                    {\mathbf{u}_{\alpha0}}^\top & {\mathbf{v}_0}^\top
                \end{array}
                \right]^\top\in\mathbb{R}^{n\alpha + nv}
                \hspace{5mm}\textrm{and}\hspace{5mm}
                \mathbf{u} = \left[
                \begin{array}{ccc}
                    {\mathbf{u}_{\alpha}}^\top & {\mathbf{u}_{\kappa}}^\top & \mathbf{v}^\top
                \end{array}
                \right]^\top\in\mathbb{R}^{n\alpha + n\kappa + nv}.
            \end{align}
            Cell corner rotations of the reference configuration are $\mathbf{u}_{\kappa 0} = \mathbf{0}$ since undeformed cell sides are assumed to be straight.


        \subsection{Transformation Matrices}
\begin{figure}[htbp]
                \begin{center}
                    \subfloat[]{
                        \includegraphics[height=0.45\textwidth]{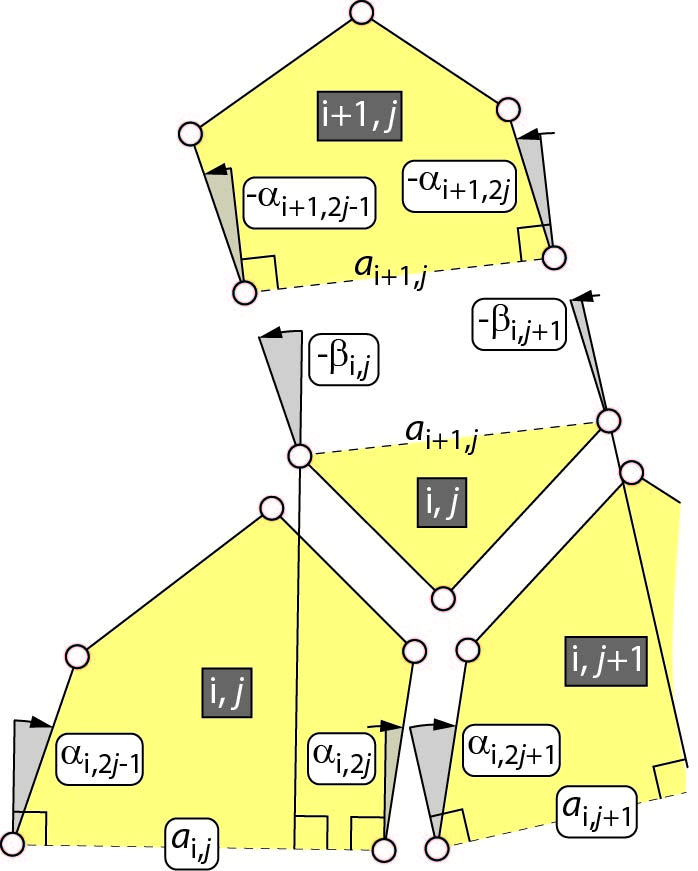}}\hspace{10mm}
                    \subfloat[]{
                        \includegraphics[height=0.45\textwidth]{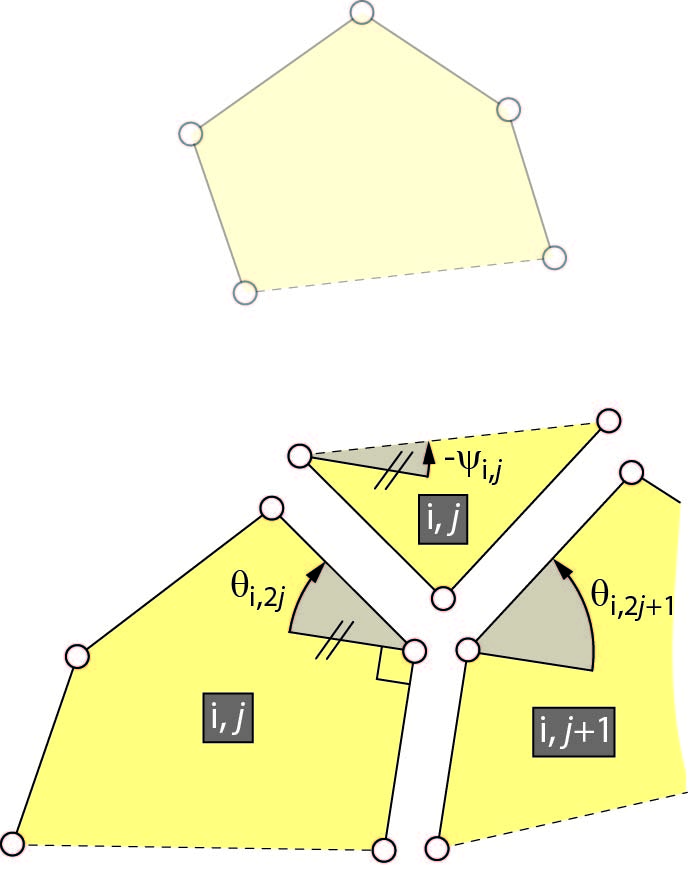}}
                    \caption{(a) State variables of pentagonal and hexagonal cells. (b) Internal angles of pentagonal and triangular cells.}
                    \label{pic:Figure_9}
                \end{center}
            \end{figure}

            As illustrated in Figure~\ref{pic:Figure_9}, state variables $\mathbf{u}^\textrm{P}_{\alpha,i+1,j}$ of the $j$-th pentagonal cell in the $(i+1)$-th cell row can be expressed in terms of state variables $\mathbf{u}^\textrm{T}_{\alpha,i,j}$, $\boldsymbol{\beta}^\textrm{T}_{i,j} =
            [\beta_{i,j}\ \ \beta_{i,j+1}]^\top$ and $\mathbf{v}^\textrm{T}_{i,j}$
            of the $j$-th triangular cell in the $i$-th cell row
            \begin{align}
                \mathbf{u}^\textrm{P}_{\alpha,i+1,j}\left(\boldsymbol{\beta}^\textrm{T}_{i,j}, \mathbf{u}^\textrm{T}_{\alpha,i,j}, \mathbf{v}^\textrm{T}_{i,j}\right) =
                \mathbf{T}^\beta_{i,j} \boldsymbol{\beta}^\textrm{T}_{i,j} + \mathbf{T}^\textrm{lin}_{i,j} \mathbf{u}_{\alpha,i,j}^\textrm{T} + \mathbf{T}^\textrm{nlin}_{i,j}\left(\mathbf{u}^\textrm{T}_{\alpha,i,j}, \mathbf{v}^\textrm{T}_{i,j}\right)
            \end{align}
            where the linear and nonlinear matrices are
            \begin{align}
                \mathbf{T}^\beta_{i,j} =
                \left[
                \begin{array}{cc}
                    1 & 0\\
                    0 & 1\\
                    0 & 0
                \end{array}
                \right]
                \textrm{,}\hspace{5mm}
                \mathbf{T}^{\textrm{lin}}_{i,j} =
                \left[
                \begin{array}{cccccc}
                    0 & -1 &  0 & 0 & 0 & 0\\
                    0 &  0 & -1 & 0 & 0 & 0\\
                    0 &  0 &  0 & 0 & 0 & 0
                \end{array}
                \right]
                \hspace{5mm}\textrm{and}\hspace{5mm}
                \mathbf{T}_{i,j}^\textrm{nlin}\left(\mathbf{u}_{\alpha,i,j}^\textrm{T},\mathbf{v}_{i,j}^\textrm{T}\right) =
                \left[
                \begin{array}{c}
                    -\psi_{i,j}\left(\mathbf{u}_{\alpha,i,j}^\textrm{T},\mathbf{v}_{i,j}^\textrm{T}\right)\\
                    -\psi_{i,j}\left(\mathbf{u}_{\alpha,i,j}^\textrm{T},\mathbf{v}_{i,j}^\textrm{T}\right)\\
                    a_{i+1,j}\left(\mathbf{u}_{\alpha,i,j}^\textrm{T},\mathbf{v}_{i,j}^\textrm{T}\right)
                \end{array}
                \right].
            \end{align}
            The transformation matrix $\mathbf{T}^{\alpha}_{i,j}$ relates pentagonal state variables $\mathbf{u}_\alpha^\textrm{P}$ to triangular state variables $\mathbf{u}_\alpha^\textrm{T}$. Similarly, the transformation matrix $\mathbf{T}^v_{i,j}$ relates pentagonal state variables $\mathbf{u}_\alpha^\textrm{P}$ to triangular state variables $\mathbf{v}^\textrm{T}$
            \begin{align}
                \mathbf{T}^{\alpha}_{i,j} =
                \frac{\partial\mathbf{u}_{\boldsymbol{\alpha},i+1,j}^\textrm{P}} {\partial\mathbf{u}_{\boldsymbol{\alpha},i,j}^\textrm{T}} =
                \mathbf{T}_{i,j}^\textrm{lin} + \frac{\partial\mathbf{T}_{i,j}^\textrm{nlin}}{\partial\mathbf{u}_{\alpha,i,j}^\textrm{T}}
                \hspace{5mm}\textrm{and}\hspace{5mm}
                \mathbf{T}^v_{i,j} =
                \frac{\partial\mathbf{u}_{\boldsymbol{\alpha},i+1,j}^\textrm{P}} {\partial\mathbf{v}_{i,j}^\textrm{T}} =
                \frac{\partial\mathbf{T}_{i,j}^\textrm{nlin}}{\partial\mathbf{v}_{i,j}^\textrm{T}}.
            \end{align}
            Transformation matrices for reference state variables are derived in a similar manner and denoted as, for example, $\mathbf{T}^{\alpha 0}_{i,j}$.


        \subsection{Potential Energy and Equilibrium Configuration}
            The potential energy of a cellular structure is the sum of the pressure energy of triangular and pentagonal cells as well as the pressure and strain energy of cell sides
            \begin{align}
                \Pi\left(\mathbf{u}_0,\mathbf{u}\right) =
                \sum_{i=1}^{nR} &\left(\ \
                \sum_{j=1}^{nP+2-i} \hspace{-2mm} \Pi_{b,i,j}^\textrm{S}\left(\mathbf{u}_0,\mathbf{u}\right)\right.\\\nonumber
                &\ \ \left.+ \hspace{-2mm}
                \sum_{j=1}^{nP+1-i} \hspace{-1mm} \left( \Pi_{i,j}^\textrm{P}\left(\mathbf{u}\right) +
                \delta_i^1 \Pi_{a,j}^\textrm{S}\left(\mathbf{u}_0,\mathbf{u}\right) + \Pi_{c,i,2j-1}^\textrm{S}\left(\mathbf{u}_0,\mathbf{u}\right) +
                \Pi_{c,i,2j}^\textrm{S}\left(\mathbf{u}_0,\mathbf{u}\right) \right)\right.\\\nonumber
                &\ \ \left.+ \left(1-\delta_i^{nR}\right) \sum_{j=1}^{nP-i} \Pi_{i,j}^\textrm{T}\left(\mathbf{u}\right) \right)
            \end{align}
            where $\delta_{i}^{nR}$ is a Kronecker delta. It can be seen that only the energies of cell sides are a function of the reference (manufactured) configuration. The gradient $\boldsymbol{\Pi}^u = \partial\Pi/\partial\mathbf{u} \in \mathbb{R}^{n\alpha+n\kappa+nv}$ of the potential energy with respect to the state variables $\mathbf{u}$ is computed by adding the contributions of single cell rows. The energy gradient $\boldsymbol{\Pi}_i^u$ incorporates only terms from cells and sides in the $i$-th cell row. It is expressed with respect to state variables $\mathbf{u}_i$ of a cellular structure that solely consists of cell rows $i \ldots n_R$
            \begin{align}
                \boldsymbol{\Pi}^u_{i} =&\hspace{-1mm}
                \sum_{j=1}^{nP+1-i}
                \left(
                \boldsymbol{\Pi}^{\textrm{P},\alpha}_{i,j}
                \frac{\partial\mathbf{u}^\textrm{P}_{\alpha,i,j}}{\partial\mathbf{u}_i}
                +
                \boldsymbol{\Pi}^{\textrm{P},v}_{i,j} \frac{\partial \mathbf{v}^\textrm{P}_{i,j}}{\partial \mathbf{u}_i}
                \right)
                +
                \left(1-\delta_i^{nR}\right)
                \sum_{j=1}^{nP-i}
                \left(
                \boldsymbol{\Pi}_{i,j}^{\textrm{T},\alpha}
                \frac{\partial\mathbf{u}^\textrm{T}_{\alpha,i,j}}{\partial\mathbf{u}_i}
                +
                \boldsymbol{\Pi}^{\textrm{T},v}_{i,j}
                \frac{\partial \mathbf{v}^\textrm{T}_{i,j}}{\partial \mathbf{u}_i}
                \right)\\\nonumber
                &\hspace{-4mm}+
                \sum_{j=1}^{nP+1-i}
                \delta_i^1
                \left(
                \boldsymbol{\Pi}^{\textrm{S},uP}_{a,i,j} \frac{\partial\mathbf{u}^\textrm{P}_{i,j}}{\partial\mathbf{u}_i} +
                \boldsymbol{\Pi}^{\textrm{S},\kappa}_{a,i,j} \frac{\partial\mathbf{u}^\textrm{S}_{\kappa,i,j}}{\partial\mathbf{u}_i} +
                \boldsymbol{\Pi}^{\textrm{S},v}_{a,i,j} \frac{\partial v^\textrm{S}_{a,i,j}}{\partial \mathbf{u}_i}
                \right)\\\nonumber
                &\hspace{-4mm}+
                \sum_{j=1}^{nP+2-i}
                \left(
                \boldsymbol{\Pi}^{\textrm{S},\kappa}_{b,i,j} \frac{\partial\mathbf{u}^\textrm{S}_{\kappa,i,j}}{\partial\mathbf{u}_i} +
                \boldsymbol{\Pi}^{\textrm{S},v}_{b,i,j}
                \frac{\partial v^\textrm{S}_{b,i,j}}{\partial\mathbf{u}_i}
                \right)\\\nonumber
                &\hspace{-4mm}+
                \sum_{j=1}^{nP+1-i}
                \left(
                \left(
                \boldsymbol{\Pi}^{\textrm{S},uP}_{c,i,2j-1}
                \frac{\partial \mathbf{u}^\textrm{P}_{i,j}}{\partial \mathbf{u}_i} +
                \boldsymbol{\Pi}^{\textrm{S},\kappa}_{c,i,2j-1} \frac{\partial\mathbf{u}^\textrm{S}_{\kappa,c,i,2j-1}}{\partial\mathbf{u}_i} +
                \boldsymbol{\Pi}^{\textrm{S},v}_{c,i,2j-1} \frac{\partial v^\textrm{S}_{c,i,2j-1}}{\partial\mathbf{u}_i}\right)
                +
                \left(
                \boldsymbol{\Pi}^{\textrm{S},uP}_{c,i,2j}
                \frac{\partial \mathbf{u}^\textrm{P}_{i,j}}{\partial \mathbf{u}_i} +
                \boldsymbol{\Pi}^{\textrm{S},\kappa}_{c,i,2j} \frac{\partial\mathbf{u}^\textrm{S}_{\kappa,c,i,2j}}{\partial\mathbf{u}_i} +
                \boldsymbol{\Pi}^{\textrm{S},v}_{c,i,2j}
                \frac{\partial v^\textrm{S}_{c,i,2j}}{\partial\mathbf{u}_i}\right)
                \right).
            \end{align}
            Terms such as $\partial\mathbf{u}_{\alpha,i,j}^\textrm{P}/\partial\mathbf{u}_i$ map state variables of single cells or sides to the state variables $\mathbf{u}_i$. Adding and transforming the gradients of single cell rows from top to bottom leads to
            \begin{align}
                \boldsymbol{\Pi}^u = \frac{\partial \Pi}{\partial \mathbf{u}} =
                \left( \left( \left( \boldsymbol{\Pi}^u_{nR} \frac{\partial\mathbf{u}_{nR}}{\partial\mathbf{u}_{nR-1}} + \ldots + \boldsymbol{\Pi}^u_{4} \right) \frac{\partial\mathbf{u}_{4}}{\partial\mathbf{u}_{3}} +
                \boldsymbol{\Pi}^u_{3} \right) \frac{\partial\mathbf{u}_{3}}{\partial\mathbf{u}_{2}} +
                \boldsymbol{\Pi}^u_{2} \right)
                \frac{\partial\mathbf{u}_{2}}{\partial\mathbf{u}_{1}} +
                \boldsymbol{\Pi}^u_{1},
            \end{align}
            where terms such as $\partial\mathbf{u}_{i+1}/\partial\mathbf{u}_i$ are assembled from previously introduced transformation matrices. The corresponding gradient with respect to reference state variables $\mathbf{u}_0$ is denoted as $\boldsymbol{\Pi}^0$. Cell pressures in CPACS are assumed to be constant throughout each cell row. Combinations of cell row pressures are subsequently referred to as pressure sets. A cellular structure is in equilibrium for a pressure set $q$ if its potential energy is stationary i.e.
            \begin{align}
                \boldsymbol{\Pi}^u_q\left(\mathbf{u}_0,\mathbf{u}_q\right) = \mathbf{0}
                \hspace{5mm}\textrm{equilibrium condition.}
            \end{align}
            This nonlinear set of equations for the state variables $\mathbf{u}_q$ can be solved by using a Newton based approach. State variables of the $(i+1)$-th Newton iteration are
            \begin{align}
                \mathbf{u}^{i+1}_q = \mathbf{u}^i_q - \left(\boldsymbol{\Pi}^{uu}_q\right)^{-1} {\boldsymbol{\Pi}^u_q}^\top
            \end{align}
            where $\boldsymbol{\Pi}^{uu}_q$ is the Hessian of the potential energy with respect to state variables $\mathbf{u}_q$.


        \subsection{Optimization}
\begin{figure}[htbp]
                \begin{center}
                    \subfloat{
                        \includegraphics[height=0.28\textwidth]{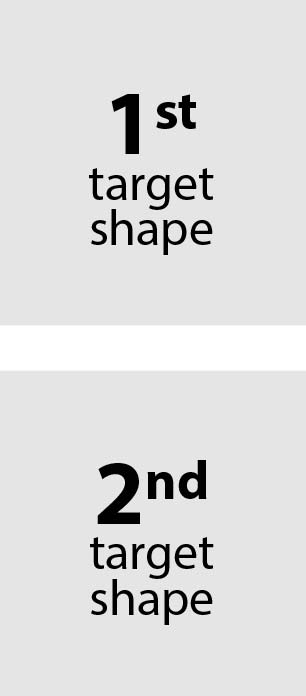}}\hspace{1mm}
                    \addtocounter{subfigure}{-1}
                    \subfloat[]{
                        \includegraphics[height=0.28\textwidth]{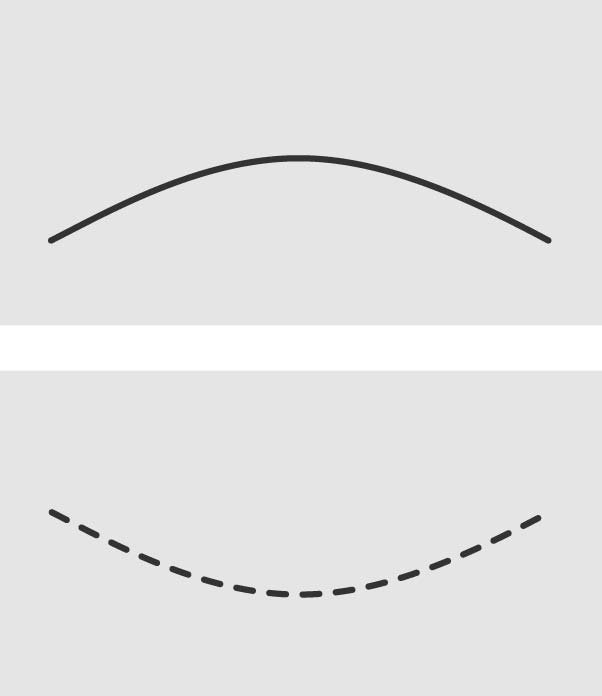}}\hspace{2mm}
                    \subfloat[]{
                        \includegraphics[height=0.28\textwidth]{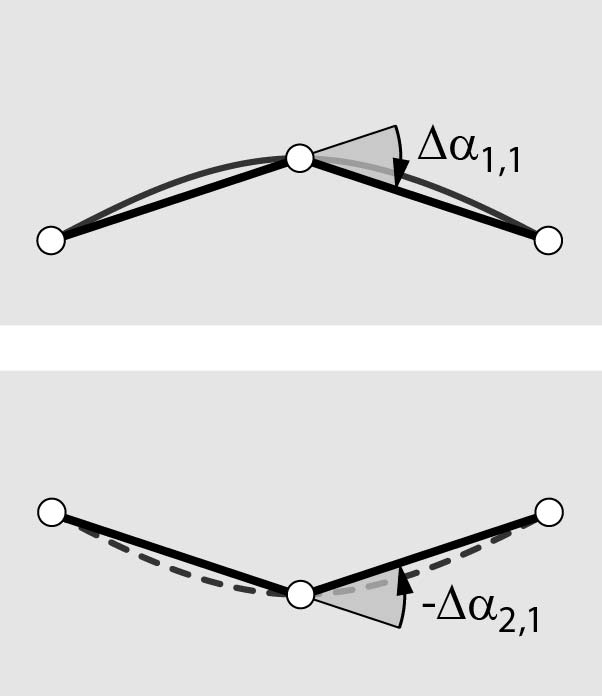}}\hspace{2mm}
                    \subfloat[]{
                        \includegraphics[height=0.28\textwidth]{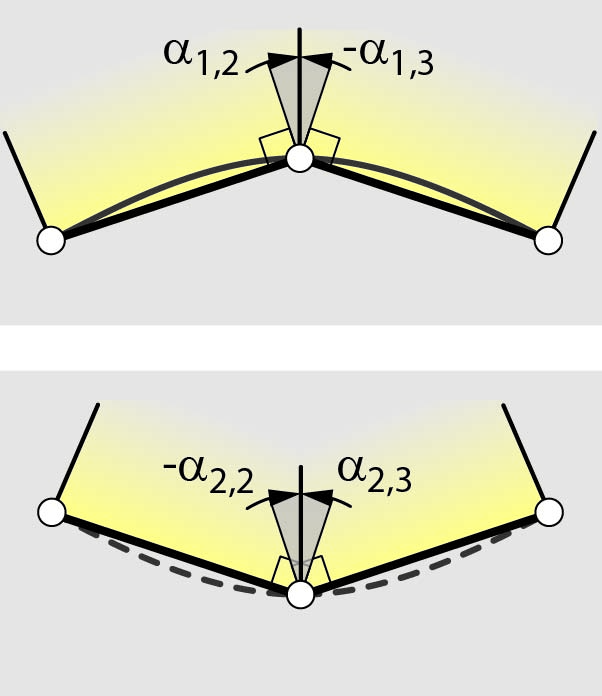}}
                    \caption{(a) Target shapes are (b) approximated by straight lines. (c) Stylized base pentagons with cell sides $\mathbf{a}$, $\mathbf{b}$ and state variables $\boldsymbol{\alpha}$.}
                  	\label{pic:Figure_10}
                \end{center}
            \end{figure}

            Side lengths of a pressure actuated cellular structure with $n_R$ cell rows can be optimized such that the outer pentagonal cell corners of an equilibrium configuration are, depending on the pressure set, located on $n_R$ different $C^1$ continuous target shapes (Figure~\ref{pic:Figure_10}). A target shape, indexed by $q$, is approximated by a piecewise linear curve with angles $\Delta \alpha_{q,j}$ at corner points. Therefore, the state variables $\boldsymbol{\alpha}_q \subset \mathbf{u}_{\alpha,q}$ of the $q$-th equilibrium configuration as shown in Figure~\ref{pic:Figure_8}(b) have to satisfy
            \begin{align}
                \Delta\alpha_{q,j} = \alpha_{q,2j}-\alpha_{q,2j+1}\hspace{5mm}\textrm{for}\hspace{5mm} j=1,\ldots,n_P-1.
            \end{align}
            In other words, a target shape is defined as a set of angles between adjacent base pentagons so that there are $n_P-1$ target angles for each equilibrium configuration. The deviation between the target shape and an equilibrium configuration at the $j$-th cell corner is given by
            \begin{align}
                r_{q,j} = \Delta \alpha_{q,j}-\alpha_{q,2j}+\alpha_{q,2j+1}
            \end{align}
            and gathered in the residual vector $\mathbf{r}_q$ for the $q$-th equilibrium configuration
            \begin{align}
                \mathbf{r}_q =
                \left[
                \begin{array}{ccc}
                    r_{q,1} &
                    \hspace{-2mm}\ldots\hspace{-2mm} &
                    r_{q,nP-1}
                \end{array}
                \right]^\top.
            \end{align}
            In turn, the residual vectors of all equilibrium configurations are gathered in
            \begin{align}
                \mathbf{r} =
                \left[
                \begin{array}{ccc}
                    {\mathbf{r}_1}^\top &
                    \hspace{-2mm}\ldots\hspace{-2mm} &
                    {\mathbf{r}_{nR}}^\top
                \end{array}
                \right]^\top.
            \end{align}
            The target angles $\Delta\alpha_{q,j}$ are a function of the base lengths $a_j$ and thus depend on the corresponding axial strains. However, their influence is neglected since these strains are usually small.
            The change of the current state variables $\mathbf{u}_q$ of the $q$-th equilibrium configuration with respect to reference state variables $\mathbf{u}_0$ is subsequently derived. Infinitesimally small variations of reference and current state variables need to satisfy
            \begin{align}
                \boldsymbol{\Pi}^u_q\left(\mathbf{u}_0 + \Delta\mathbf{u}_0, \mathbf{u}_q + \Delta\mathbf{u}_q\right) = \mathbf{0}.
            \end{align}
            Neglecting higher order terms leads to
            \begin{align}
                \boldsymbol{\Pi}^u_q + \boldsymbol{\Pi}^{u0}_q \Delta\mathbf{u}_0 + \boldsymbol{\Pi}^{uu}_q \Delta\mathbf{u}_q = \mathbf{0}
            \end{align}
            where $\boldsymbol{\Pi}^u_q = \mathbf{0}$ from which we can deduce the gradient
            \begin{align}
                \mathbf{G}_q =
                \frac{\partial\mathbf{u}_q}{\partial\mathbf{u}_0} =
                -\left(\boldsymbol{\Pi}^{uu}_q\right)^{-1} {\boldsymbol{\Pi}^{0u}_q}^\top.
            \end{align}
            The matrix
            \begin{align}
                \mathbf{H} = \frac{\partial\mathbf{r}}{\partial\mathbf{u}_0} =
                \left[
                \begin{array}{ccc}
                    \displaystyle
                    {\frac{\partial\mathbf{r}_1}{\partial\mathbf{u}_0}}^\top
                    \hspace{-2mm}\ldots\hspace{-2mm} &
                    \displaystyle
                    {\frac{\partial\mathbf{r}_{nR}}{\partial\mathbf{u}_0}}^\top
                \end{array}
                \right]^\top =
                \left[
                \begin{array}{ccc}
                    {\mathbf{G}_1}^\top\mathbf{B}^\top &
                    \hspace{-2mm}\ldots\hspace{-2mm} &
                    {\mathbf{G}_{nR}}^\top\mathbf{B}^\top
                \end{array}
                \right]^\top\in\mathbb{R}^{nR\left(nP-1\right) \times \left(n\alpha+nv\right)}
            \end{align}
            relates the residual vector $\mathbf{r}$ to reference state variables $\mathbf{u}_0$ where $\mathbf{B}$ is a Boolean matrix. It is not quadratic and therefore not invertible. Hence, there exists a null-space $\mathbf{N} = \textrm{null}\left(\mathbf{H}\right)$ with $\dim\left(\mathbf{N}\right) = n_\alpha+n_v-n_R\left(n_P-1\right)$ where changes in state variables $\mathbf{u}_0$ do not affect the residual vector $\mathbf{r}$. In other words, it is possible to minimize an arbitrary objective function within the null-space \cite{Rega2005} where $\mathbf{r} = \mathbf{0}$. For example, the objective function $\mathcal{F}$ can be chosen as
            \begin{align}
                \mathcal{F}\left(\boldsymbol{\mathfrak{u}}_0\right) = \frac{1}{2} \left(\boldsymbol{\mathfrak{u}}_0 - \boldsymbol{\mathfrak{u}}_t\right)^\top \left(\boldsymbol{\mathfrak{u}}_0 - \boldsymbol{\mathfrak{u}}_t\right)
            \end{align}
            which minimizes the difference between state variables $\boldsymbol{\mathfrak{u}}_0$ and target values $\boldsymbol{\mathfrak{u}}_t$. In the following, it is assumed that $\boldsymbol{\mathfrak{u}}_0 = \{\mathbf{v}_0\}\backslash\{\mathbf{a}_0\}$ so that only reference cell side lengths other than the pentagonal base sides are varied during the optimization. The gradient and Hessian of $\mathcal{F}$ are
            \begin{align}
                \boldsymbol{\mathcal{F}}^0 = \frac{\partial \mathcal{F}}{\partial\boldsymbol{\mathfrak{u}}_0} = \boldsymbol{\mathfrak{u}}_0 - \boldsymbol{\mathfrak{u}}_t
                \hspace{5mm}\textrm{and}\hspace{5mm}
                \boldsymbol{\mathcal{F}}^{00} = \frac{\partial^2 \mathcal{F}}{{\partial\boldsymbol{\mathfrak{u}}_0}^2} = \mathbf{I},
            \end{align}
            where $\mathbf{I}$ is an identity matrix of size $n_v - n_P$. Therefore, the optimization problem can be stated as
            \begin{align}
                &\textrm{minimize }  \hspace{3mm} \mathcal{F}\left(\boldsymbol{\mathfrak{u}}_0\right)\\\nonumber
                &\textrm{subject to }\hspace{3mm} \mathbf{r} = \mathbf{0}
            \end{align}
            which can be solved with Lagrange multipliers and the Newton method. The Lagrangian
            \begin{align}
                \mathcal{L}\left(\boldsymbol{\mathfrak{u}}_0,\boldsymbol{\lambda}\right) = \mathcal{F}\left(\boldsymbol{\mathfrak{u}}_0\right) + \boldsymbol{\lambda}^\top \mathbf{r}\left(\boldsymbol{\mathfrak{u}}_0\right)
            \end{align}
            is stationary if
            \begin{align}
                \frac{\partial \mathcal{L}}{\partial \boldsymbol{\mathfrak{u}}_0} = \mathbf{0}
                \hspace{5mm}\textrm{and}\hspace{5mm}
                \frac{\partial \mathcal{L}}{\partial \boldsymbol{\lambda}} = \mathbf{0}.
            \end{align}
            The set of nonlinear equations for $\boldsymbol{\mathfrak{u}}_0$ and $\boldsymbol{\lambda}$ is iteratively solved with the Newton method which leads to
            \begin{align}
                \left[
                \begin{array}{cc}
                    \mathbf{I} + \boldsymbol{\mathcal{Z}}^i & {\boldsymbol{\mathcal{H}}^i}^\top\\
                    \boldsymbol{\mathcal{H}}^i & \mathbf{0}
                \end{array}
                \right]
                \left[
                \begin{array}{c}
                    \boldsymbol{\mathfrak{u}}_0^{i+1} - \boldsymbol{\mathfrak{u}}_0^i\\
                    \boldsymbol{\lambda}^{i+1} - \boldsymbol{\lambda}^i
                \end{array}
                \right] = -
                \left[
                \begin{array}{c}
                    \boldsymbol{\mathcal{F}}^{0,i} + {\boldsymbol{\mathcal{H}}^i}^\top \boldsymbol{\lambda}^i\\
                    \mathbf{r}^i
                \end{array}
                \right]
                \hspace{5mm}\textrm{where}\hspace{5mm}
                \boldsymbol{\mathcal{H}} = \frac{\partial \mathbf{r}}{\partial \boldsymbol{\mathfrak{u}}_0}
                \textrm{,}\hspace{5mm}
                \mathcal{Z}_{i,j} = \sum_{k=1}^{nR\left(nP-1\right)} \lambda_k \frac{\partial\mathcal{H}_{k,i}}{\partial\mathfrak{u}_{0,j}}.
            \end{align}
            Computing the nonlinear contributions of the constraint equation requires third-order derivatives in $\boldsymbol{\mathcal{Z}}$ which are computationally expensive. On the other hand, neglecting these terms can slow down convergence. This problem can be overcome by sacrificing the objective function. If the target values are dynamically chosen at each iteration such that
            \begin{align}
                \boldsymbol{\mathfrak{u}}_0^i - \boldsymbol{\mathfrak{u}}_t^i = \mathbf{0}
            \end{align}
            then the Newton method reduces to
            \begin{align}
                \left[
                \begin{array}{cc}
                    \mathbf{I} & {\boldsymbol{\mathcal{H}}^i}^\top\\
                    \displaystyle \boldsymbol{\mathcal{H}}^i & \mathbf{0}
                \end{array}
                \right]
                \left[
                \begin{array}{c}
                    \boldsymbol{\mathfrak{u}}_0^{i+1} - \boldsymbol{\mathfrak{u}}_0^i\\
                    \boldsymbol{\lambda}^{i+1} - \boldsymbol{\lambda}^i
                \end{array}
                \right] = -
                \left[
                \begin{array}{c}
                    \mathbf{0}\\
                    \mathbf{r}^i
                \end{array}
                \right]
            \end{align}
            since $\boldsymbol{\mathcal{F}}^{0,i} = \mathbf{0}$ and therefore $\boldsymbol{\lambda}^i = \mathbf{0}$.


    \section{Example Structures}
\begin{figure}[htbp]
            \centering
            \begin{minipage}[c]{0.23\textwidth}
                \subfloat[]{
                    \includegraphics[width=1\linewidth]{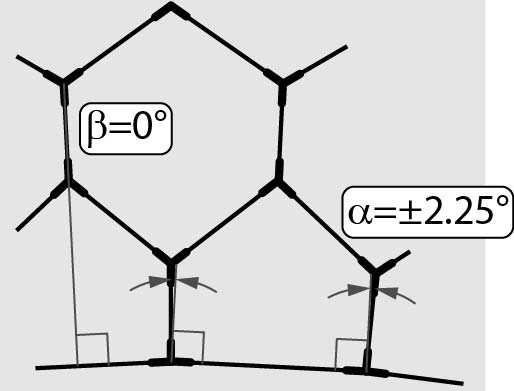}}\vspace{5mm}

                \subfloat[]{
                    \includegraphics[width=1\linewidth]{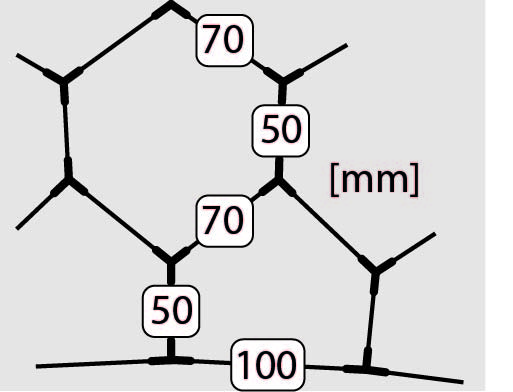}}
            \end{minipage}\hspace{12mm}
            \begin{minipage}[c]{0.6\textwidth}
                \subfloat[]{
                    \includegraphics[width=1\linewidth]{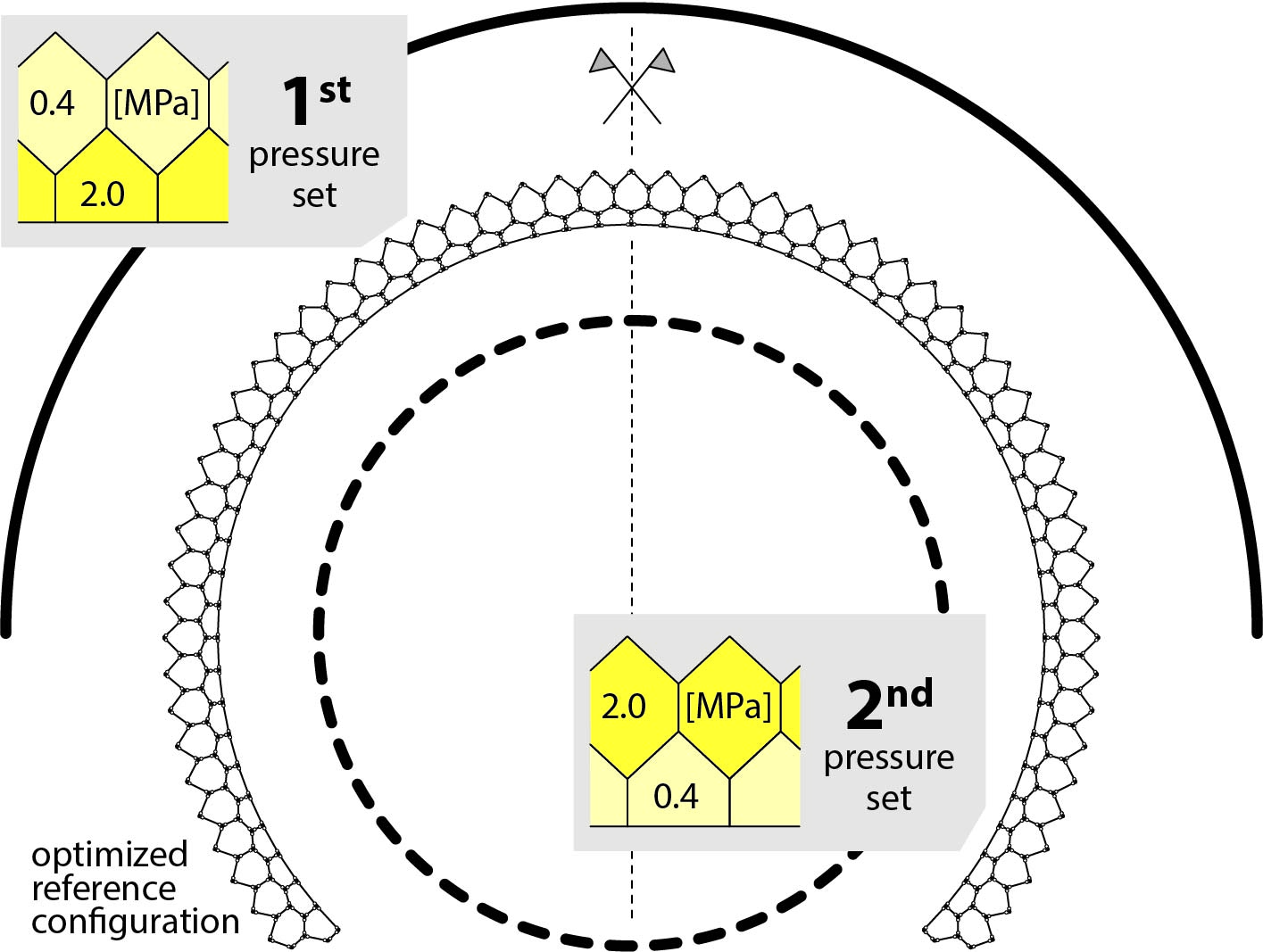}}
            \end{minipage}
            \caption{Reference state variables (a) $\mathbf{u}_{\alpha 0}$ and (b) initial cell side lengths $\mathbf{v}_0$ (c) Target shapes with associated pressure sets and optimized reference configuration.}
            \label{pic:Figure_11}
        \end{figure}

\begin{figure}[htbp]
            \begin{center}
                \subfloat[]{
                    \includegraphics[height=0.45\textwidth]{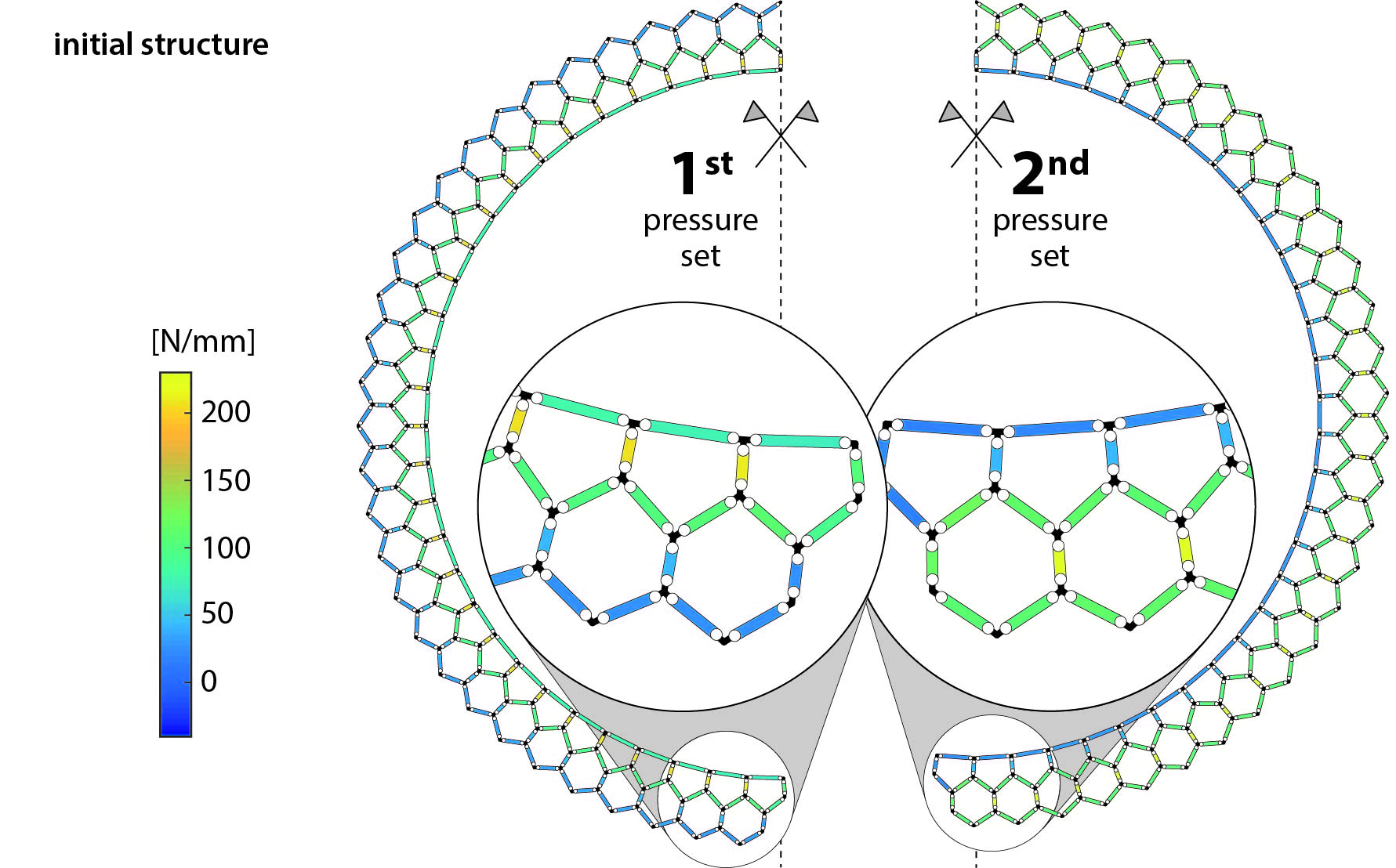}}\hspace{5mm}
                \subfloat[]{
                    \includegraphics[height=0.45\textwidth]{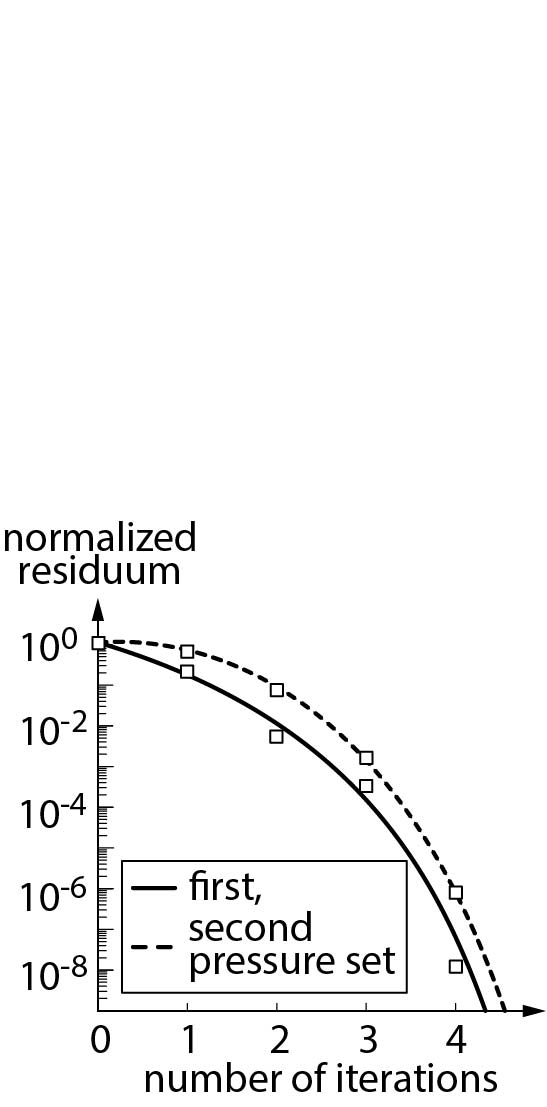}}

                \subfloat[]{
                    \includegraphics[height=0.45\textwidth]{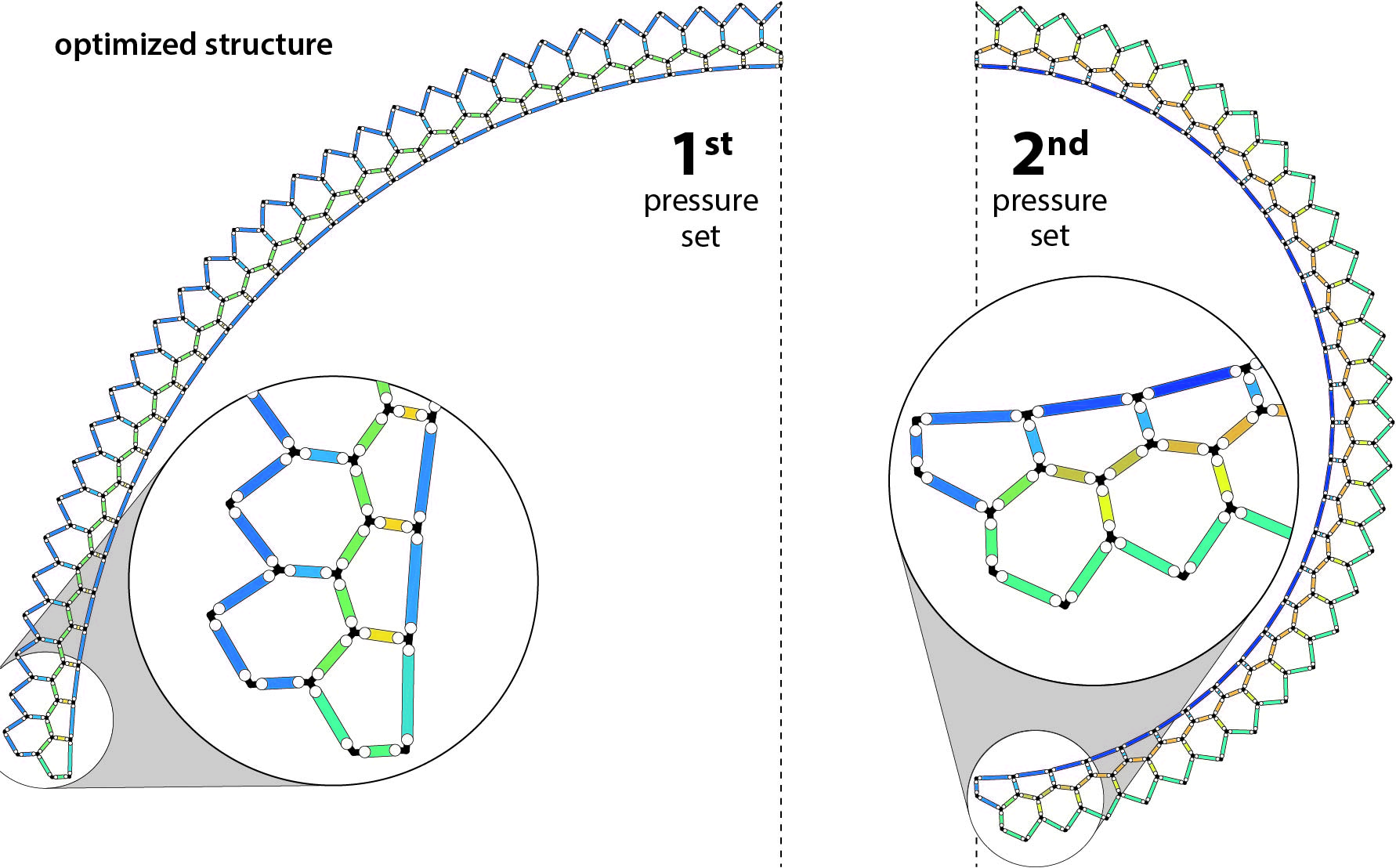}}\hspace{5mm}
                \subfloat[]{
                    \includegraphics[height=0.45\textwidth]{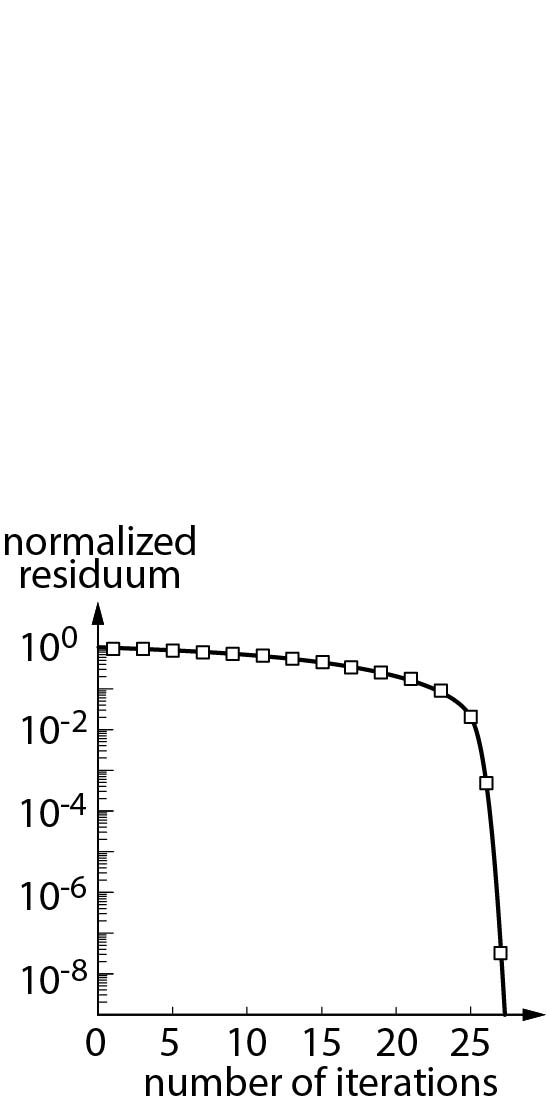}}
                \caption{Equilibrium shapes of the (a) initial and (c) optimized structure for both pressure sets. Convergence plots for computing the (b) equilibrium shapes of the initial structure and (d) optimal cell side lengths.}
                \label{pic:Figure_12}
            \end{center}
        \end{figure}

\begin{figure}[htbp]
            \begin{center}
                \begin{minipage}[c]{0.2\textwidth}
                    \subfloat[]{
                        \includegraphics[width=1\textwidth]{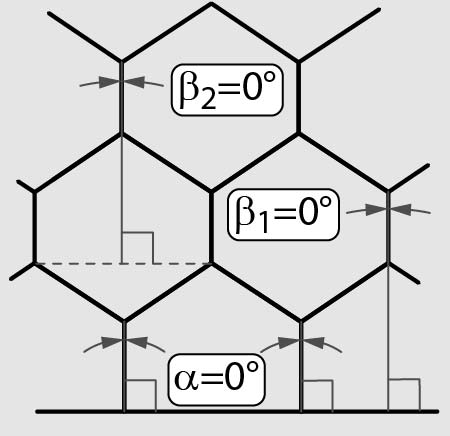}}\vspace{5mm}

                    \subfloat[]{
                        \includegraphics[width=1\textwidth]{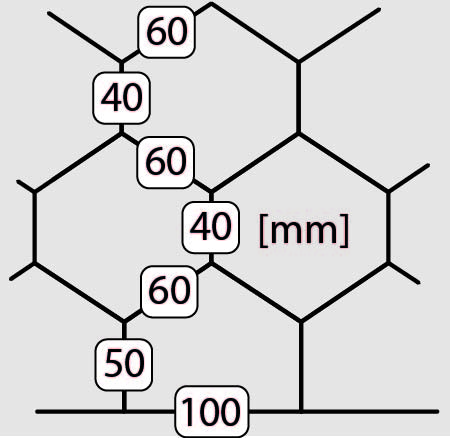}}\vspace{5mm}

                    \subfloat[]{
                        \includegraphics[width=1\textwidth]{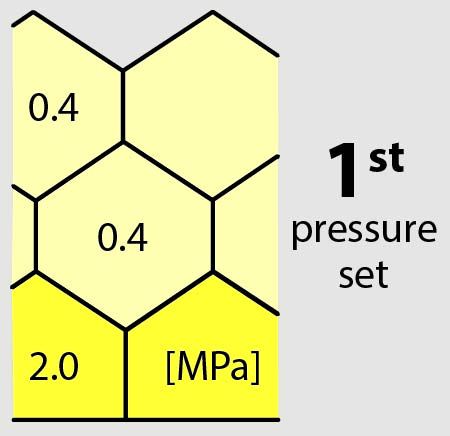}}\vspace{5mm}

                    \subfloat[]{
                        \includegraphics[width=1\textwidth]{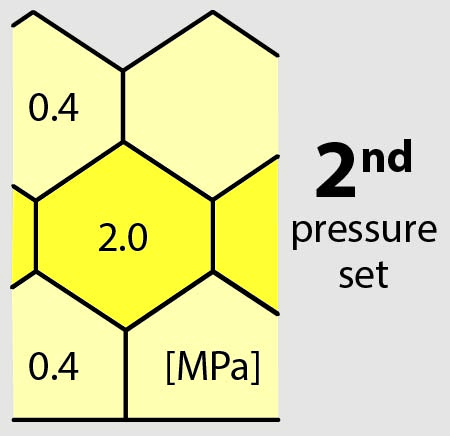}}\vspace{5mm}

                    \subfloat[]{
                        \includegraphics[width=1\textwidth]{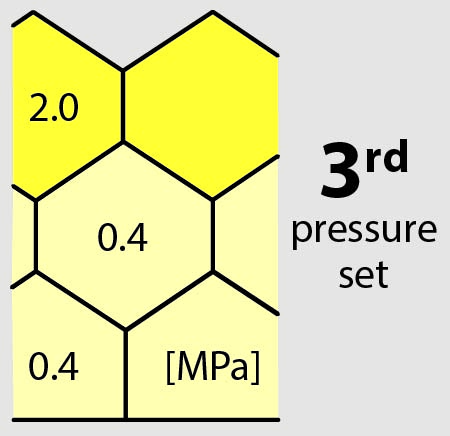}}
                \end{minipage}\hspace{10mm}
                \begin{minipage}[c]{0.7\textwidth}
                    \subfloat[]{
                        \includegraphics[height=1.86\textwidth]{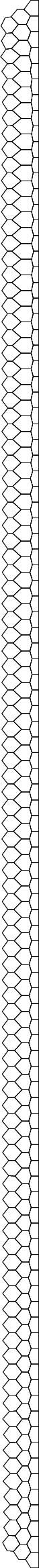}}
                    \subfloat[]{
                        \includegraphics[height=1.86\textwidth]{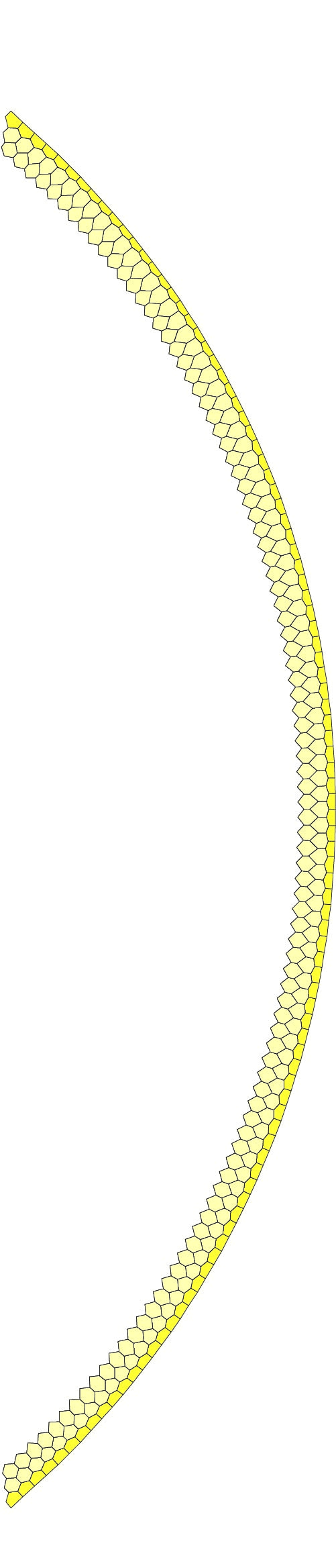}}
                    \subfloat[]{
                        \includegraphics[height=1.86\textwidth]{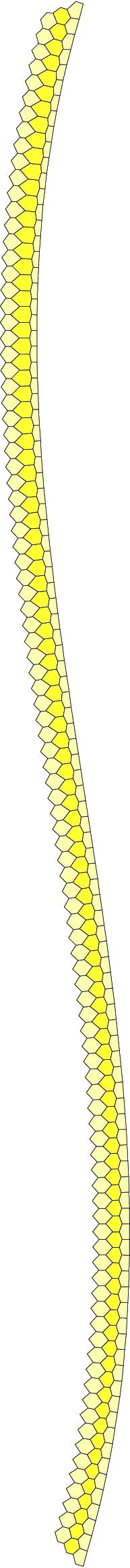}}                        \subfloat[]{
                        \includegraphics[height=1.86\textwidth]{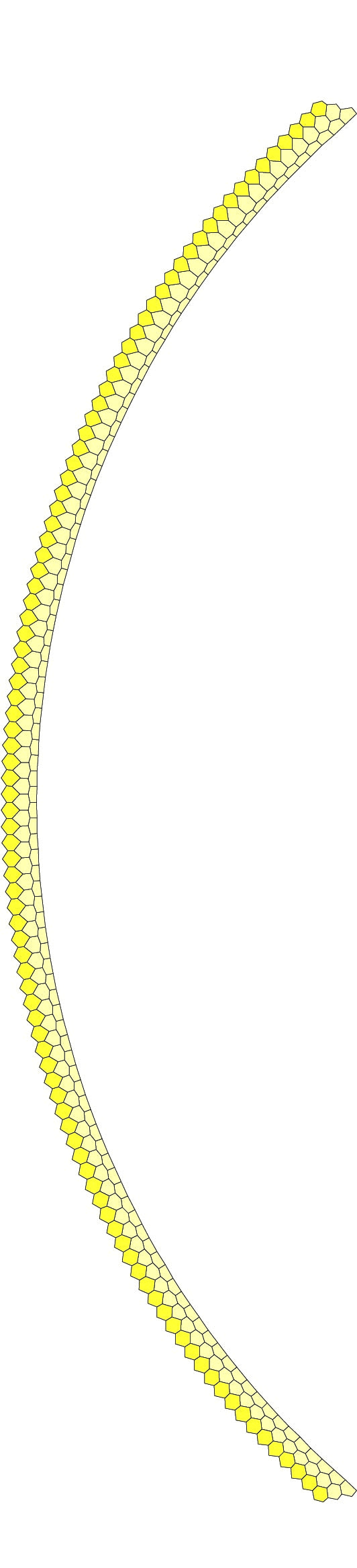}}
                \end{minipage}
                \caption{(a) Reference state variables $\mathbf{u}_{\alpha 0}$, (b) initial cell side lengths $\mathbf{v}_0$ and (c-e) three different pressure sets. (f) Optimized reference configuration and (g-i) corresponding equilibrium configurations for all pressure sets.}
              	\label{pic:Figure_13}
            \end{center}
        \end{figure}

        An example structure (Figure~\ref{pic:Figure_11}) that consists of two cell rows with 60 pentagonal and 59 hexagonal cells is used to demonstrate the performance of the proposed algorithm. The first target shape is a full circle and the second target shape is a half circle. Furthermore, the left and right boundaries are not constrained. Therefore, boundary cells change their shape due to pressure variations. The presented results are based on a complete structural simulation and optimization so that 359 cell side lengths are optimized. Irrespective of the boundary conditions, edge effects cause varying lengths along the structure so that it is not possible to solve this problem by investigating only a few cells. The hinge eccentricities $d$, rotational- $e$ and axial springs $h$ for a structure with a unit depth are chosen as
        \begin{align}\nonumber
            d = 10~\textrm{mm}
            \textrm{,}\hspace{5mm}
            e = \frac{3}{2}~\textrm{kN}
            \hspace{5mm}\textrm{and}\hspace{5mm}
            h = 3~\textrm{kN/mm}^2.
        \end{align}
        The equilibrium shapes and axial cell side forces of the original and optimized structure as well as the convergence plots for computing the equilibrium shapes and the optimal cell side lengths are shown in Figure~\ref{pic:Figure_12} for both pressure sets. It can be seen that the equilibrium configurations of the optimized structure reassemble a half- and full circle. Furthermore, optimized cell side lengths differ significantly from the initial structure and vary between both ends.
        Convergence plots show that the equilibrium configurations are computed in four iterations whereas the shape optimization requires 27 iterations. The presented optimization results are based on the second optimization approach that avoids the computation of third-order derivatives by sacrificing the objective function. This decision is motivated by the fact that, for this example, the results from the constrained optimization are very similar to the results from the second optimization approach. The presented results are computed with the help of Matlab on a single i5-4250U CPU core at 1.3 GHz in about 5 minutes. The optimized reference configuration and corresponding equilibrium shapes of a second example structure that consists of three cell rows with 100 pentagonal and 197 hexagonal cells is shown in Figure~\ref{pic:Figure_13}. It can be seen that the proposed framework is capable of optimizing cellular structures for more than two target shapes and a varying curvature. The chosen hinge eccentricities $d$, rotational- $e$ and axial springs $h$ for a structure with a unit depth are
        \begin{align}\nonumber
            d = 0
            \textrm{,}\hspace{5mm}
            e = \frac{25}{12}~\textrm{kN}
            \hspace{5mm}\textrm{and}\hspace{5mm}
            h = \infty.
        \end{align}
        The target angles at the left and right boundary are set to zero for all pressure sets
        \begin{align}
            \Delta \alpha_{i,j} = 0
            \hspace{3mm}\textrm{for}\hspace{3mm}
            j \in \left\{1,2,n_P-2,n_P-1\right\}
        \end{align}
        since the shape changing capability of the structure at both ends is reduced. Therefore, cell geometries at both ends are invariant to pressure variations. The three target shapes are
        \begin{align}
            \Delta \alpha_{1,j}  = -1^\circ,\hspace{3mm}
            \Delta \alpha_{2,j}  = \frac{j-3}{n_P-6} 2^\circ - 1^\circ
            \hspace{5mm}\textrm{and}\hspace{5mm}
            \Delta \alpha_{3,j}  = 1^\circ
            \hspace{3mm}\textrm{for}\hspace{3mm}
            j \in \left\{3,\ldots,n_P-3\right\}.
        \end{align}


    \section{Conclusions}
        This article presents a novel approach for the simulation and optimization of compliant pressure actuated cellular structures (CPACS). It complements previous work \cite{Pagitz2012-1} by taking into account an arbitrary number of cell rows, rotational/axial springs and hinge eccentricities at cell corners. The kinematic framework splits naturally into two parts. The first part describes pressure actuated cellular structures with central cell corner hinges. The second part adds a rotational degree of freedom at each cell corner to describe the state of nonzero hinge eccentricities. Furthermore, it has been shown that the geometric primitives of CPACS are triangular and pentagonal cells as well as cell sides. This geometric reduction can be used as a basis for an object oriented implementation. The convergence rate of the optimization is, compared to previous work \cite{Pagitz2012-1}, significantly enhanced by using a Newton method. This will enable the computation of the sensitivity of an optimal solution with respect to hinge eccentricities and rotational/axial springs in future work. Hence it is possible to directly use the presented framework for the dimensioning of a cellular structure. The presented approach will ultimately be the basis for a software tool that can directly send the optimization results to a rapid prototyping machine.



    \newpage

    \begin{appendices}
        \small
        \sectionfont{\large}
        \subsectionfont{\normalsize}
        \section{Appendix}
            \subsection{Pentagonal Cells}
            \renewcommand\theequation{\thesection.\arabic{equation}}
            \setcounter{equation}{0}
            Derivatives of the internal variables and pressure potential of a pentagonal cell with respect to state variables $\mathbf{u}_\alpha^\textrm{P}$, $\mathbf{v}^\textrm{P}$ are subsequently summarized. Derivatives of length $y$ are
            \begin{align}\label{eqn:dy_duP}
                \frac{\partial y}{\partial \mathbf{u}^\textrm{P}_{\alpha}} =
                \frac{1}{y}
                \left[
                \begin{array}{c}
                    \ \ \ \left(\sin\left(\alpha_{12}\right)b_2 - a\cos\left(\alpha_1\right)\right)b_1\\
                    -\left(\sin\left(\alpha_{12}\right)b_1 - a\cos\left(\alpha_2\right)\right)b_2\\
                    y_x
                \end{array}
                \right]^\top
                \textrm{,}\hspace{3mm}
                \frac{\partial y}{\partial \mathbf{v}^\textrm{P}} =
                \frac{1}{y}
                \left[
                \begin{array}{c}
                    -\sin\left(\alpha_1\right)a - \cos\left(\alpha_{12}\right)b_2 + b_1\\
                    \ \ \ \sin\left(\alpha_2\right)a - \cos\left(\alpha_{12}\right)b_1 + b_2\\
                    0\\
                    0
                \end{array}
                \right]^\top
                \hspace{3mm}\textrm{where}\hspace{3mm}
                \alpha_{12} = \alpha_1-\alpha_2.
            \end{align}
            Derivatives of altitude $z$ are
            \begin{align}
                \begin{split}
                \frac{\partial z}{\partial \mathbf{u}^\textrm{P}_{\alpha}} &=
                \frac{\partial z}{\partial y}\frac{\partial y}{\partial \mathbf{u}^\textrm{P}_{\alpha}}\\
                \frac{\partial z}{\partial \mathbf{v}^\textrm{P}} &=
                \frac{\partial z}{\partial y}\frac{\partial y}{\partial \mathbf{v}^\textrm{P}} + \frac{\partial z}{\partial \mathbf{c}^\textrm{P}} \frac{\partial\mathbf{c}^\textrm{P}}{\partial\mathbf{v}^\textrm{P}}
                \end{split}
                \hspace{10mm}\textrm{where}\hspace{5mm}
                \begin{split}
                \frac{\partial z}{\partial y} &= \frac{1}{4y^3z}\left(\left({c_1}^2 - {c_2}^2\right)^2 - y^4\right)\\
                \frac{\partial z}{\partial \mathbf{c}^\textrm{P}} &= \frac{1}{2y^2z}
                \left[
                \begin{array}{cc}
                    c_1\left(y^2-{c_1}^2+{c_2}^2\right) & c_2\left(y^2+{c_1}^2-{c_2}^2\right)
                \end{array}
                \right].
                \end{split}
            \end{align}
            Expressions for the internal angle $\theta_1$ are
            \begin{align}
                \begin{split}
                    \displaystyle\frac{\partial \theta_1}{\partial \mathbf{u}^\textrm{P}_{\alpha}} &=
                    \frac{1}{y_x}
                    \left(\frac{\partial y_y}{\partial \mathbf{u}^\textrm{P}_{\alpha}} -
                    \frac{y_y}{y} \frac{\partial y}{\partial \mathbf{u}^\textrm{P}_{\alpha}}\right) + \frac{\rho_1}{\sqrt{{c_1}^2 - z^2}}\frac{\partial z}{\partial \mathbf{u}^\textrm{P}_{\alpha}} +
                    \left[
                    \begin{array}{ccc}
                        1 & 0 & 0
                    \end{array}
                    \right]\\
                    \displaystyle\frac{\partial \theta_1}{\partial \mathbf{v}^\textrm{P}} &=
                    \frac{1}{y_x}
                    \left(\frac{\partial y_y}{\partial \mathbf{v}^\textrm{P}} -
                    \frac{y_y}{y} \frac{\partial y}{\partial \mathbf{v}^\textrm{P}}\right) +
                    \frac{\rho_1}{\sqrt{{c_1}^2 - z^2}}\left(\frac{\partial z}{\partial \mathbf{v}^\textrm{P}}
                    - \frac{z}{c_1}
                    \left[
                    \begin{array}{cccc}
                        0 & 0 & 1 & 0
                    \end{array}
                    \right]\right)
                \end{split}
                \hspace{5mm}\textrm{where}\hspace{5mm}
                \begin{split}
                    \rho_1 =
                    \begin{cases}
                        \hspace{2mm} 1 & {c_2}^2 \leq y^2 + {c_1}^2\\
                        -1 & {c_2}^2 > y^2 + {c_1}^2
                    \end{cases}
                \end{split}
            \end{align}
            \noindent and the gradients of the pressure potential $\Pi^\textrm{P}$ are
            \begin{align}\label{eqn:dP_duP}
                \boldsymbol{\Pi}^{\textrm{P},\alpha} &=
                -\frac{p}{2} \left(
                \left[
                \begin{array}{c}
                    -\left(\sin\left(\alpha_1\right) a + \cos\left(\alpha_{12}\right) b_2\right) b_1\\
                    -\left(\sin\left(\alpha_2\right) a - \cos\left(\alpha_{12}\right) b_1\right) b_2\\
                    \cos\left(\alpha_1\right) b_1 + \cos\left(\alpha_2\right) b_2
                \end{array}
                \right]^\top +
                z \frac{\partial y}{\partial \mathbf{u}^\textrm{P}_{\alpha}} +
                y \frac{\partial z}{\partial \mathbf{u}^\textrm{P}_{\alpha}}\right)
                \\\nonumber
                \boldsymbol{\Pi}^{\textrm{P},v} &=
                -\frac{p}{2} \left(
                \left[
                \begin{array}{c}
                    \cos\left(\alpha_1\right) a^\textrm{P} - \sin\left(\alpha_{12}\right) b_2\\
                    \cos\left(\alpha_2\right) a^\textrm{P} - \sin\left(\alpha_{12}\right) b_1\\
                    0\\
                    0
                \end{array}
                \right]^\top +
                z \frac{\partial y}{\partial \mathbf{v}^\textrm{P}} +
                y \frac{\partial z}{\partial \mathbf{v}^\textrm{P}}\right).
            \end{align}


        \subsection{Triangular Cells}
            Derivatives of the internal variables and pressure potential of a triangular cell with respect to state variables $\mathbf{u}_\alpha^\textrm{T}$, $\mathbf{v}^\textrm{T}$ are subsequently summarized. Derivatives of abstract side $a$ are
            \begin{align}\label{eqn:da_duT}
                \frac{\partial a}{\partial \mathbf{u}^\textrm{T}_{\alpha}} &=
                -\frac{1}{a} c_2 c_3 \sin\left(\theta_1+\theta_2\right) \frac{\partial \left(\theta_1+\theta_2\right)}{\partial \mathbf{u}^\textrm{T}_{\alpha}}\\\nonumber
                \frac{\partial a}{\partial \mathbf{v}^\textrm{T}} &=
                -\frac{1}{a} \left(
                c_2 c_3 \sin\left(\theta_1+\theta_2\right) \frac{\partial\left(\theta_1 + \theta_2\right)}{\partial\mathbf{v}^\textrm{T}}
                -\left[
                \begin{array}{ccccccc}
                    0 & 0 & 0 & 0 & c_2 + c_3 \cos\left(\theta_1+\theta_2\right) & c_3 + c_2 \cos\left(\theta_1+\theta_2\right) & 0
                \end{array}
                \right]\right).
            \end{align}
            Derivatives of angle $\psi$ are
            \begin{align}\label{eqn:dpsi_duT}
                \frac{\partial \psi}{\partial \mathbf{u}^\textrm{T}_{\alpha}} =
                \frac{\partial \theta_1}{\partial \mathbf{u}^\textrm{T}_{\alpha}}
                + \frac{\partial \psi}{\partial a} \frac{\partial a}{\partial \mathbf{u}^\textrm{T}_{\alpha}}
                \textrm{,}\hspace{5mm}
                \frac{\partial \psi}{\partial \mathbf{v}^\textrm{T}} =
                \frac{\partial \theta_1}{\partial \mathbf{v}^\textrm{T}}
                + \frac{\partial \psi}{\partial a} \frac{\partial a}{\partial \mathbf{v}^\textrm{T}} + \frac{\partial \psi}{\partial \mathbf{c}^\textrm{T}} \frac{\partial\mathbf{c}^\textrm{T}}{\partial\mathbf{v}^\textrm{T}}
            \end{align}
            where derivatives with respect to abstract base side $a$ and cell sides $\mathbf{c}^\textrm{T} = \left[
            \begin{array}{cccc}
                c_1 & c_2 & c_3 & c_4
            \end{array}
            \right]^\top$ are
            \begin{align}\label{eqn:dpsi_da}
                \frac{\partial \psi}{\partial a} =
                \frac{{a}^2 - {c_2}^2 + {c_3}^2}{a^\textrm{T} \sqrt{\left(2 a c_2\right)^2 - \left({a}^2 + {c_2}^2 - {c_3}^2\right)^2}}
                \textrm{,}\hspace{5mm}
                \frac{\partial \psi}{\partial \mathbf{c}^\textrm{T}} =
                -\frac{1}{c_2 \sqrt{\left(2 a c_2\right)^2 - \left({a}^2 + {c_2}^2 - {c_3}^2\right)^2}}
                \left[
                \begin{array}{c}
                    0\\
                    {a}^2 - {c_2}^2 - {c_3}^2\\
                    2 c_2 c_3\\
                    0
                \end{array}
                \right]^\top.
            \end{align}
            The gradient of the pressure potential is
            \begin{align}\label{eqn:dP_duT}
                \boldsymbol{\Pi}^{\textrm{T},\alpha} &= -\frac{p}{2} c_2 c_3 \cos\left(\theta_1+\theta_2\right)
                \frac{\partial\left(\theta_1+\theta_2\right)}{\partial\mathbf{u}^\textrm{T}_{\alpha}}\\\nonumber
                \boldsymbol{\Pi}^{\textrm{T},v} &= -\frac{p}{2} \left( c_2 c_3 \cos\left(\theta_1+\theta_2\right) \frac{\partial\left(\theta_1+\theta_2\right)}{\partial\mathbf{v}^\textrm{T}} + \sin\left(\theta_1+\theta_2\right)
                \left[
                \begin{array}{ccccccc}
                    0 & 0 & 0 & 0 & c_3^\textrm{T} & c_2^\textrm{T} & 0
                \end{array}
                \right] \right).
            \end{align}


        \subsection{Cell Sides}
            Derivatives of the internal variables and pressure, strain potential of a cell side with respect to state variables $\mathbf{u}^\textrm{S}_{\kappa}$, $v^\textrm{S}$ are subsequently summarized. Derivatives of the length $L_h$ between cell side hinges are
            \begin{alignat}{2}\label{eqn:dL_duS}
                &\frac{\partial L_h}{\partial \mathbf{u}^\textrm{S}_{\kappa}} &&=
                \frac{1}{L_h}\left(L_{hx} \frac{\partial L_{hx}}{\partial \mathbf{u}_\kappa^\textrm{S}} + L_{hy} \frac{\partial L_{hy}}{\partial \mathbf{u}_\kappa^\textrm{S}}\right) =
                \frac{1}{L_h}\left(
                L_{hx}
                \left[
                \begin{array}{cc}
                    \sin\left(\kappa_-\right) d_- &
                    \sin\left(\kappa_+\right) d_+
                \end{array}
                \right]
                +
                L_{hy}
                \left[
                \begin{array}{cc}
                    \cos\left(\kappa_-\right) d_- &
                    \cos\left(\kappa_+\right) d_+
                \end{array}
                \right]
                \right)\\\nonumber
                &\frac{\partial L_h}{\partial v^\textrm{S}} &&=
                \frac{L_{hx}}{L_h}.
            \end{alignat}
            Derivatives of, for example, the bending angle $\varphi_-$ are
            \begin{align}\label{eqn:dphi_duS}
                \frac{\partial \varphi_-}{\partial \mathbf{u}_{\kappa}^\textrm{S}} =
                \left[
                \begin{array}{cc}
                    1 &
                    0
                \end{array}
                \right]
                +
                \frac{1}{L_{hx}}
                \left(
                \frac{\partial L_{hy}}{\partial \mathbf{u}_\kappa^\textrm{S}}
                -
                \frac{L_{hy}}{L_h}
                \frac{\partial L_h}{\partial \mathbf{u}_\kappa^\textrm{S}}
                \right)
                \hspace{5mm}\textrm{and}\hspace{5mm}
                \frac{\partial \varphi_-}{\partial v^\textrm{S}} =
                -\frac{L_{hy}}{L_{hx} L_h}
                \frac{\partial L_h}{\partial v^\textrm{S}}.
            \end{align}
            Gradients of the pressure and strain potential are
            \begin{alignat}{2}\label{eqn:dP_duSp}
                &\boldsymbol{\Pi}^{\textrm{S},\kappa}_p &&=
                - \frac{\Delta p}{2}\left(
                \left[
                \begin{array}{c}
                    \left(\sin\left(\kappa_-\right)^2
                    - \cos\left(\kappa_-\right)^2\right) {d_-}^2
                    - \cos\left(\kappa_-\right) d_- L_{hx}\\
                    \left(\cos\left(\kappa_+\right)^2
                    - \sin\left(\kappa_+\right)^2\right) {d_+}^2
                    +\cos\left(\kappa_+\right) d_+ L_{hx}
                \end{array}
                \right]
                +
                \left(
                \sin\left(\kappa_+\right) d_+ -
                \sin\left(\kappa_-\right) d_-
                \right)
                \frac{\partial L_{hx}}{\partial \mathbf{u}_{\kappa}^\textrm{S}}\right)\\\nonumber
                &\boldsymbol{\Pi}^{\textrm{S},v}_p &&=
                -\frac{\Delta p}{2} \left( \sin\left(\kappa_+\right) d_+ - \sin\left(\kappa_-\right) d_-\right) \frac{\partial L_{hx}}{\partial v^\textrm{S}}
            \end{alignat}
            and
            \begin{align}\label{dP_duSe}
                \boldsymbol{\Pi}^{\textrm{S},\kappa}_e =
                e_- \varphi_- \frac{\partial \varphi_-}{\partial \mathbf{u}_\kappa^\textrm{S}} +
                e_+ \varphi_+ \frac{\partial \varphi_+}{\partial \mathbf{u}_\kappa^\textrm{S}} + h \Delta L_h \frac{\partial L_h}{\partial \mathbf{u}_\kappa^\textrm{S}}
                \textrm{,}\hspace{5mm}
                \boldsymbol{\Pi}^{\textrm{S},v0}_e =
                -h \Delta L_h
                \hspace{5mm}\textrm{and}\hspace{5mm}
                \boldsymbol{\Pi}^{\textrm{S},v}_e =
                e_- \varphi_- \frac{\partial \varphi_-}{\partial v^\textrm{S}} +
                e_+ \varphi_+ \frac{\partial \varphi_+}{\partial v^\textrm{S}} + h \Delta L_h \frac{\partial L_h}{\partial v^\textrm{S}}.
            \end{align}
            The gradient of the total cell side energy with respect to pentagonal state variables
            $\mathbf{u}^\textrm{P} = \left[\mathbf{u}_\alpha^\textrm{P} \hspace{2mm} \mathbf{v}^\textrm{P}\right]^\top$ is
            \begin{align}\label{eqn:dP_duS}
                \boldsymbol{\Pi}^{\textrm{S},uP} =
                \boldsymbol{\Pi}^{\textrm{S},\kappa}
                \left(
                \frac{\partial \mathbf{u}_{\kappa}^\textrm{S}}{\partial \mathbf{u}_{\alpha}^\textrm{P}}
                \frac{\partial \mathbf{u}_{\alpha}^\textrm{P}}{\partial\mathbf{u}^\textrm{P}}
                +
                \frac{\partial \mathbf{u}_{\kappa}^\textrm{S}}{\partial \boldsymbol{\beta}^\textrm{P}}
                \frac{\partial \boldsymbol{\beta}^\textrm{P}}{\partial\mathbf{u}^\textrm{P}}
                +
                \frac{\partial \mathbf{u}_\kappa^\textrm{S}}{\partial\mathbf{v}^\textrm{P}} \frac{\partial\mathbf{v}^\textrm{P}}{\partial\mathbf{u}^\textrm{P}}
                \right).
            \end{align}
    \end{appendices}
\end{document}